\documentclass[aps,prd,twocolumn,nofootinbib,floatfix,preprintnumbers]{revtex4}
\usepackage{subfigure}
\usepackage{graphicx}
\usepackage{dcolumn}
\usepackage{epsfig}
\usepackage{amsmath}
\usepackage{amsfonts}
\usepackage{amssymb}
\usepackage{color}
\usepackage{hyperref}

\newcommand{\rev}[1]{\textcolor{black}{{}{#1}{}}}

\newcommand{\be}{\begin{equation}}
\newcommand{\ee}{\end{equation}}
\newcommand{\bea}{\begin{eqnarray}}
\newcommand{\eea}{\end{eqnarray}}

\begin{document}

\title{Laguerre reconstruction of the BAO feature in halo-based mock galaxy catalogues}

\author{Farnik Nikakhtar}
\email{farnik@sas.upenn.edu}
\affiliation{Department of Physics and Astronomy, University of Pennsylvania, Philadelphia, PA 19104 -- USA}




\author{Ravi K.~Sheth}
\affiliation{Center for Particle Cosmology, University of Pennsylvania, Philadelphia, PA 19104 -- USA}
\affiliation{The Abdus Salam International Center for Theoretical Physics, Strada Costiera 11, Trieste 34151 -- Italy}

\author{Idit Zehavi}
\affiliation{Department of Physics, Case Western Reserve University, Cleveland, OH 44106-7079 -- USA}

\date{\today}

\begin{abstract}
  Fitting half-integer generalized Laguerre functions to the evolved, real-space dark matter and halo correlation functions provides a simple way to reconstruct their initial shapes.  We show that this methodology also works well in a wide variety of realistic, assembly biased, velocity biased and redshift-space distorted mock galaxy catalogs.  We use the {\em linear point} feature in the monopole of the redshift-space distorted correlation function to quantify the accuracy of our approach. We find that the linear point estimated from the mock galaxy catalogs is insensitive to the details of the biasing scheme at the subpercent level.  However, the linear point scale in the nonlinear, biased, and redshift-space distorted field is systematically offset from its scale in the unbiased linear density fluctuation field by more than 1\%.  In the Laguerre reconstructed correlation function, this is reduced to sub-percent values, so it provides comparable accuracy and precision to methods that reconstruct the full density field before estimating the distance scale.  The linear point in the reconstructed density fields provided by these other methods is likewise precise, accurate, and insensitive to galaxy bias.  All reconstructions depend on some input parameters, and marginalizing over uncertainties in the input parameters required for reconstruction can degrade both accuracy and precision.  The linear point simplifies the marginalization process, enabling more realistic estimates of the precision of the distance scale estimate for negligible additional computational cost.  We show this explicitly for Laguerre reconstruction.
\end{abstract}


\maketitle

\newcommand{\ste}[1]{\textcolor{red}{\textbf{\small[Ste: #1]}}}


\section{Introduction}\label{intro}
The Baryon Acoustic Oscillation feature in the galaxy distribution can be used as a standard cosmological ruler \cite{peeblesYu,esw2007,sanchezDR12,alamDR12}.  In practice, there are several details associated with defining this ruler, some of which make more explicit use of the expected shape of the feature than others.  Gravitational evolution modifies the shape of the feature, potentially biasing and degrading the cosmological constraints which current and future datasets can provide.  This has driven the development of many algorithms which seek to restore the feature to its original shape \cite{esss07, recPW, recSDSS, kn2017, recIterate, recHE, eFAM2019, royaMAK}.

Recently, we proposed a simple reconstruction algorithm \cite{LPlaguerre} which makes minimal use of the expected shape of the feature, but instead exploits the fact that, to leading order, the evolved feature is related to the original one by a convolution \cite{Bharadwaj1996,rpt}.  The convolution is approximately an isotropic Gaussian, so this motivated the fitting of a set of (integer or half-integer) generalized Laguerre functions to the measured two-point correlation function of galaxies.  We used the Linear Point feature in this clustering signal \cite{PaperI, PRDmocks, PRLboss, LPnus, LPruler} to quantify the gains in accuracy and precision which result from Laguerre reconstruction of the correlation functions of dark matter, low and high mass halos, finding that they are comparable to those returned by more traditional and sophisticated algorithms which seek to reconstruct the density field rather than just its two-point statistics.

While promising, there were two respects in which our previous tests were incomplete.  The first is that they were performed using measurements in real space.  Observations are made in redshift space, and the associated distortions with respect to real space \cite{kaiser87, hamilton92} can be thought of as arising from additional convolutions \cite{fisher95, rsdPeaks}.  This extra smearing may impact the quality of the Laguerre reconstruction.  In addition, there is not a one-to-one relationship between galaxies and dark matter halos:  rather, galaxies are complex tracers of the dark matter distribution.  At subpercent precision, this galaxy bias has the potential to also bias cosmological constraints, especially if this bias couples to galaxy velocities and hence to redshift space distortions.

The physics which determines galaxy bias are not known a priori, so this has driven the development of mock catalogs which allow one to rapidly explore a range of galaxy bias prescriptions and their impact on the BAO signal \cite{baoHOD}. The main goal of the present paper is to study if Laguerre reconstruction, which is relatively agnostic about the background cosmological model, can be similarly agnostic about galaxy bias, even when starting from the redshift-space distorted signal.

We will continue to use the Linear Point (LP) feature to quantify the fidelity of our reconstructions.  The LP is defined as the midpoint between the peak and dip values in the two-point correlation function:  
\begin{equation}
 r_{\rm LP}\equiv \frac{r_{\rm peak} + r_{\rm dip}}{2}.  
 \label{eq:rLP}
\end{equation}
Ref.\cite{PaperI} provided an analytic argument for why, to approximately percent-level precision, the LP should be the same in real and redshift space, for all biased tracers and all times.  (The argument is easier to see for the inflection point which lies in between $r_{\rm peak}$ and $r_{\rm dip}$, so we also show results for $r_{\rm infl}$ in what follows.  In practice, $r_{\rm LP}$ turns out to be slightly more stable.)  However, at subpercent precision, there are hints that the estimated LP scale may depend slightly but systematically on halo mass.  Therefore, a second but distinct goal of our work is to use a variety of realistic, redshift-space distorted mock galaxy catalogs to test the robustness of the LP-based approach itself.  There are two distinct parts to this goal:  one is to check if the estimated distance scale is unbiased, and the second is to quantify the associated uncertainties on the estimated scale.  As we noted in Ref.~\cite{LPlaguerre}, the input parameters required for reconstruction are not known perfectly, and accounting for this will degrade the precision of the estimated scale; the LP feature simplifies the process of determining more realistic error bars.

Section~\ref{sec:motivation} motivates why Laguerre reconstruction of the monopole of the redshift space correlation function can be performed similarly to real space and describes the mock catalogs we use to illustrate our results.  Section~\ref{sec:methods} presents our measurements of the LP scale in the measured and Laguerre-reconstructed correlation functions and compares them with results from more traditional, density field reconstruction methods.  It then illustrates the degradation in precision which results from marginalizing over the values of the parameters required as input for reconstruction.  A final section summarizes.  \rev{Both Laguerre reconstruction and the LP methodology are agnostic about the (in principle unknown) shape of the dark matter correlation function, whereas the expected shape plays a key role in the method used by the Baryon Oscillation Spectroscopic Survey (BOSS) collaboration\footnote{\url{https://www.sdss3.org/surveys/boss.php}}.  An Appendix discusses and contrasts the precision and accuracy of distance scale constraints which are derived from the LP with those which come from fitting the measured correlation function to a template shape.}

\section{Motivation}\label{sec:motivation}
This section uses measurements in simulations to motivate the use of the Laguerre algorithm for reconstructing the monopole of the redshift space correlation function.  

\subsection{Effective smearing scale}
Following \cite{rpt,PaperI} we approximate the evolved redshift space monopole on BAO scales as
\begin{align}
  \xi_{\rm NL}(s) &= \int_{-1}^1 \frac{d\mu}{2}\int \frac{dk \,k^2}{2\pi^2}\,
  P_{\rm Lin}(k)\,j_0(ks)\,\Bigl[b_{10} + f\mu^2\Bigr]^2\,
  \nonumber\\
  &\qquad \qquad \times    e^{-k^2\Sigma^2(1-\mu^2)}
  e^{-k^2\Sigma^2\,(1+f)^2 \,\mu^2} \nonumber\\
  &\qquad \qquad + {\rm MC~terms},
  \label{eq:xis}
\end{align}
with $\Sigma^2 = \int dk\,P_{\rm Lin}(k,z)/3\pi^2\propto D^2(z)$ and 
$f \equiv d\ln D/d\ln a$ where $D(z)$ is the linear theory growth factor.  The real-space expression which motivated the Laguerre method has $f=0$, so the question is if the integral over $\mu$ is a serious complication.

To see that it is not, note that the integral over $\mu$ equals $b_{10}^2$ times 
\begin{align}
 &\frac{\sqrt{\pi}\, {\rm erf}(\alpha)}{2\alpha}\left(1 + \frac{\beta}{\alpha^2} + \frac{3\beta^2}{4\alpha^4}\right)
 - e^{-\alpha^2} \left(\frac{\beta}{\alpha^2} + \frac{3\beta^2}{4\alpha^4} + \frac{\beta^2}{2\alpha^2} \right)\nonumber\\
 &\to \left(1 + \frac{2\beta}{3} + \frac{\beta^2}{5}\right) -
 \frac{\alpha^2}{3} \left(1 + \frac{6\beta}{5} + \frac{3\beta^2}{7}\right)
  + \cdots ,
\end{align}
where $\alpha^2 \equiv k^2\Sigma^2 \, f(2+f)$ and $\beta \equiv f/b_{10}$.  This suggests that
\begin{equation}
  \xi(s) \approx 
    \int \frac{dk \,k^2}{2\pi^2}\,b_{\rm eff}^2P_{\rm Lin}(k,z)\,e^{-k^2\Sigma_{\rm eff}^2(z)}\,j_0(ks),
  \label{eq:xiapprox}
\end{equation}
plus the mode-coupling terms, where
\begin{align}
  b_{\rm eff}^2 &= b_{10}^2\, \left(1 + \frac{2\beta}{3} + \frac{\beta^2}{5}\right) \\
  \Sigma_{\rm eff}^2 &= \Sigma^2 \left[1 + \frac{f(2+f)}{3}\frac{1 + 6\beta/5 + 3\beta^2/7}{1 + 2\beta/3 + \beta^2/5} \right] \nonumber\\
  & \approx 1.9\,\Sigma^2  \qquad {\rm for} \qquad (f,b_{10}) = (0.75,1).
  \label{eq:sigeff}
\end{align}
Note that these rescalings depend on $f$ and $b_{10}$, but not on the shape of $P(k)$.  In addition, the factor $1.9\to 1.8$ when $b_{10}=2.1$, so the dependence on $b_{10}$ is weak.

The analysis above shows that, with the replacements $b_{10}\to b_{\rm eff}$ and $\Sigma\to\Sigma_{\rm eff}$, equation~(\ref{eq:xiapprox}) has the same form as its real space counterpart.  If we assume that the same is true of the mode-coupling term, then one should be able to reconstruct the redshift space monopole using the same Laguerre algorithm as was used for the real space correlation function.  However, because $\Sigma_{\rm eff}> \Sigma$, we expect some degradation in the constraining power of the reconstruction.  We test this expectation in the remainder of this paper.

\subsection{Simulation set}
Our analysis uses mock catalogues based on generalized halo occupation distribution (HOD) populations of 20 periodic boxes from the {\tt ABACUS COSMOS} release, the same ones that were studied by \cite[][hereafter DE2019]{baoHOD}.  (Strictly speaking, DE2019 used an additional 16 boxes from the matched `Emulator' simulation set.  We discuss why we do not use these in the Appendix.)  Each box is $1100h^{-1}$Mpc (comoving) on a side, and the background cosmology is a flat $\Lambda$CDM model with $(\Omega_{\rm cdm}h^2,\Omega_b h^2) = (0.1199,0.02222)$, and $(h,n_s,\sigma_8) = (0.6726,0.9652,0.83)$.  The associated linear theory values of $r_{\rm LP}$ and $r_{\rm infl}$ for the dark matter are $93h^{-1}$Mpc and $93.4h^{-1}$Mpc, respectively.  For easy comparison with DE2019 and Ref.\cite{LPlaguerre} we focus on the $z=0.5$ outputs for which $\Sigma$ (see text immediately following equation~\ref{eq:xis}) is $4.6 h^{-1}$Mpc. 

The number density of galaxies in all the mock catalogs is within 0.1 percent of $4\times 10^{-4} h^3$Mpc$^{-3}$.  The (monopole of the) redshift-space distorted correlation functions measured in these HOD mock catalogs were kindly made available by DE2019.  However, the measurements were made in bins of width $5h^{-1}$Mpc, which is wider than the $3h^{-1}$Mpc bins which are necessary for Laguerre reconstruction \cite{LPlaguerre}.  Therefore, we made our own measurements of the correlation functions in these mocks and checked that they were consistent with those of DE2019.  Except for this difference in bin size, our analysis is based on exactly the same HODs and (monopole) measurements as theirs.

Figure~6 of DE2019 shows how the correlation functions change as the galaxy assignment scheme (the HOD) is varied.  The different HODs produce large-scale bias factors $b_{10}$ \rev{(estimated from the ratio of the nonlinear real space power spectra of the galaxies to that of the dark matter on scales $k\le 0.1h$/Mpc)}, which can differ by of order 20\% from a fiducial value of about 2.15.  Inserting these values in equation~(\ref{eq:sigeff}) yields the effective bias factor of each redshift-space monopole, $b_{\rm eff}\approx 2.43$, and the effective smearing scale, $\Sigma_{\rm eff}\approx 6.2h^{-1}$Mpc.  Although we have checked all the HOD models, we have chosen to only present results from the two sets of paired HODs which gave the most discrepant results in DE2019 (see their Table~2 and Figure~7).

For `Base 2' and `3', the central galaxy in a halo is at rest with respect to the halo center of mass, and the models differ only in how the number of satellites scales with halo mass:  $\langle N_{\rm sat}|M\rangle\propto [(M-M_{\rm cut})/M_1]^\alpha$ with $M_{\rm cut}=10^{13.35}h^{-1}M_\odot$ and $(\alpha,M_1)=(0.75,10^{13.770}h^{-1}M_\odot)$ and $(1.25,10^{13.848}h^{-1}M_\odot)$, respectively, where the value of $M_1$ is chosen to keep the total (central + satellite) number density fixed at $4\times 10^{-4}$($h^{-1}$Mpc)$^{-3}$.  For this pair, the small scale clustering, fingers of god, and large scale bias can be different.  Both `Velocity (Cen 20\%)' and `Velocity (Cen 100\%)' have $(\alpha,M_1)=(1,10^{13.8}h^{-1}M_\odot)$, but differ in how each central galaxy moves with respect to its halo center:  the rms speed of the central is 20\%  (realistic) or 100\% (extreme) of that of the dark matter particles, respectively, potentially affecting the redshift space clustering signal.

\begin{figure}
 \centering
 \includegraphics[width=0.9\hsize]{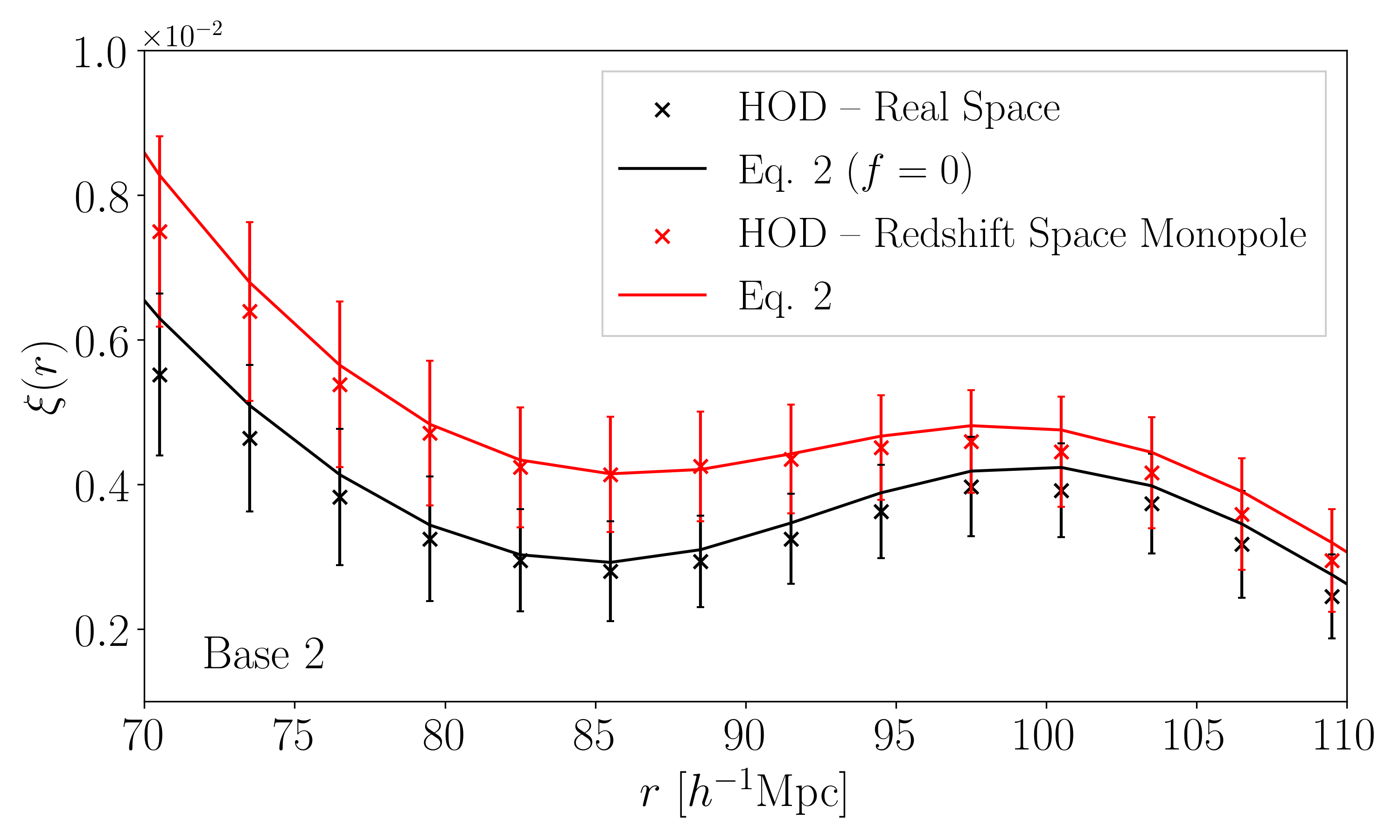}
 \caption{Real and redshift space (monopole) correlation functions in the `Base 2' HOD model of the Abacus simulation set.  Solid curves show Eq.(\ref{eq:xis}) with $\Sigma=4.6h^{-1}$Mpc, $b_{10}=2.17$, and $f=0$ (real-space) or $0.76$ (redshift space monopole) respectively.  
  }   
 \label{fig:xis}
\end{figure}

Before we consider the reconstruction of these correlation functions, we must check if Eq.~(\ref{eq:xis}) provides a reasonable description of the redshift-space monopole.  For this, we have used one of the HODs:  the black and red symbols with error bars in Figure~\ref{fig:xis} show the measured real and redshift-space correlation functions (averaged over the 20 simulation boxes).  The associated curves show Eq.~(\ref{eq:xis}) with $(\Sigma,b_{10})=(4.6h^{-1}{\rm Mpc},2.17)$ and $f=0$ or $0.76$ respectively.  Evidently, Eq.(\ref{eq:xis}) is slightly worse for the redshift space monopole.  This is also true for the other DE2019 HODs.  Nevertheless, the level of agreement suggests that Laguerre reconstruction of the monopole should still be useful.

\section{Methods and Results}\label{sec:methods}
Laguerre reconstruction, like density field reconstruction methods, requires some prior information.  For this reason, we present our analysis in two steps.  The first assumes that this information is known perfectly.  In the second step, we discuss how to proceed if one must marginalize over the uncertainties in this required prior information.  

In what follows, we use the mock catalogs in two different ways.  We either treat each realization individually, or we average together all the correlation functions for a given HOD in all the simulations and work with this average.  The former lets us quantify the effects of cosmic variance, and explore the sensitivity of the reconstruction to noisy data.  The latter is like measuring the correlation function in an effective comoving volume of $20\times (1.1h^{-1}$Gpc)$^3 \approx 27\, h^{-3}$Gpc$^3$.

\begin{table*}
  \begin{center}
    \begin{tabular}{c||c|c||c|c||c|c||c|c}
    \hline
    Tracer & $b_{\rm eff}$ & $\chi^2_{\rm dof}$ & $r_{\rm LP-pre}$ & $r_{\rm LP-rec}$ & $r_{\rm infl-pre}$ & $r_{\rm infl-rec}$ & $r_{\rm LP-stan}$ & $r_{\rm peak-stan}$\\  
    \hline
    Base 2 & 2.34 & 0.89 & 91.62 $\pm$ 0.41 & 92.98 $\pm$ 0.46 & 92.08 $\pm$ 0.43 & 93.38 $\pm$ 0.48 & 92.83 $\pm$ 0.42 & 100.56 $\pm$ 0.43 \\
    Base 3 & 2.52 & 1.10 & 91.33 $\pm$ 0.46 & 92.88 $\pm$ 0.51 & 91.81 $\pm$ 0.46 & 93.33 $\pm$ 0.52 & 92.88 $\pm$ 0.44 & 100.34 $\pm$ 0.47 \\
    Vel 20 & 2.41 & 0.69 & 91.56 $\pm$ 0.44 & 92.79 $\pm$ 0.42 & 91.99 $\pm$ 0.47 & 93.27 $\pm$ 0.46 & 92.79 $\pm$ 0.44 & 100.22 $\pm$ 0.47 \\
    Vel 100 & 2.43 & 0.78 & 91.45 $\pm$ 0.39 & 92.85 $\pm$ 0.41 & 91.92 $\pm$ 0.43 & 93.30 $\pm$  0.45 & 92.85 $\pm$ 0.42 & 100.43 $\pm$ 0.48 \\
    \hline
    \end{tabular}
  \end{center}
  \caption{Linear point and inflection scales (in $h^{-1}$Mpc) in the pre- and post-reconstruction correlation functions measured in an effective comoving volume of nearly 27~$h^{-3}$Gpc$^3$, estimated by fitting 9th order Laguerre-based functions to the $z=0.5$ two-point correlation functions (bins of width $3h^{-1}$Mpc over the range 60-120$h^{-1}$Mpc).  First two columns show the effective large scale bias and goodness of fit for each sample, and final two columns show the LP and peak scales in the `standard' reconstruction. }
  \label{tab:results}
\end{table*}

\begin{figure*}
 \centering
 \includegraphics[width=0.45\hsize]{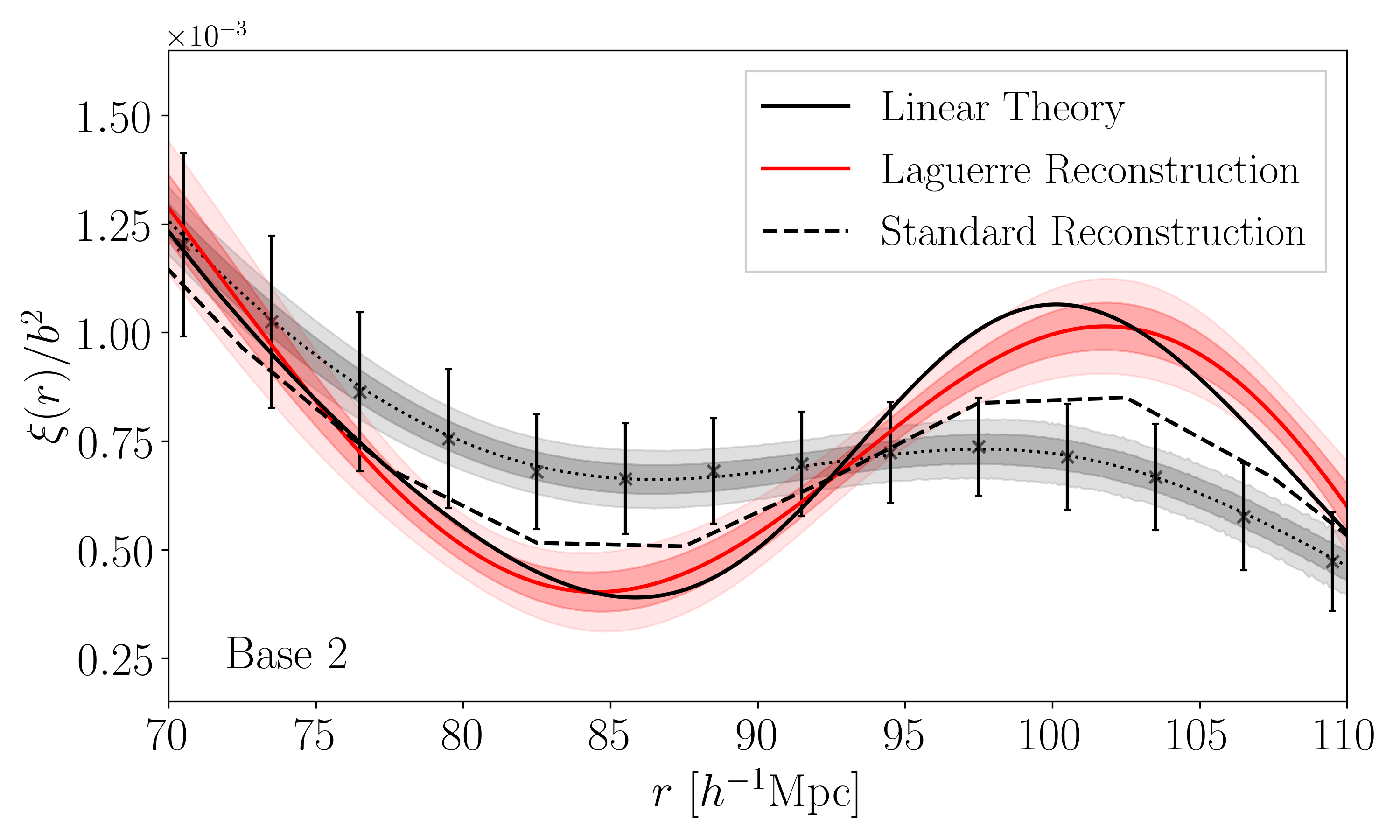}
 \includegraphics[width=0.45\hsize]{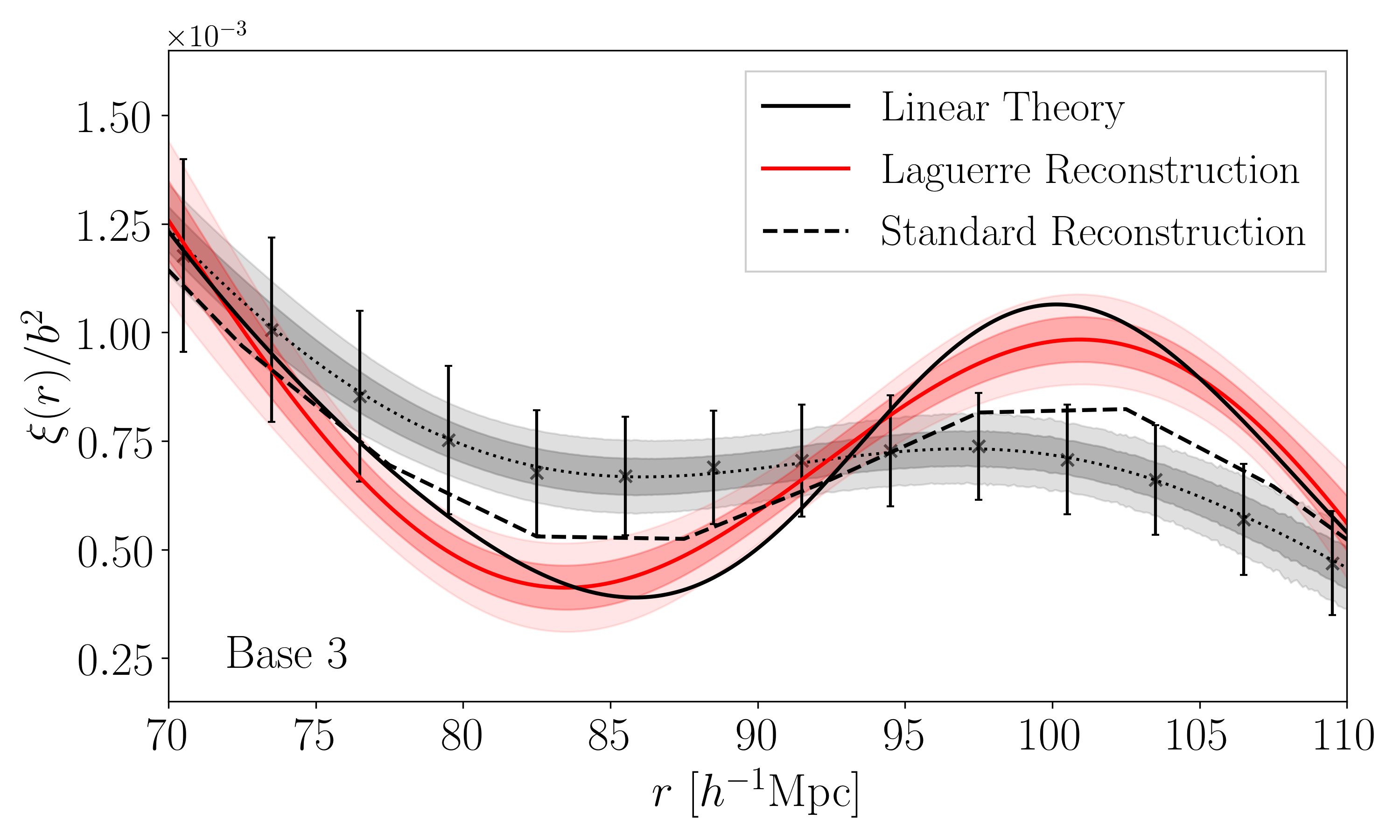}
 \includegraphics[width=0.45\hsize]{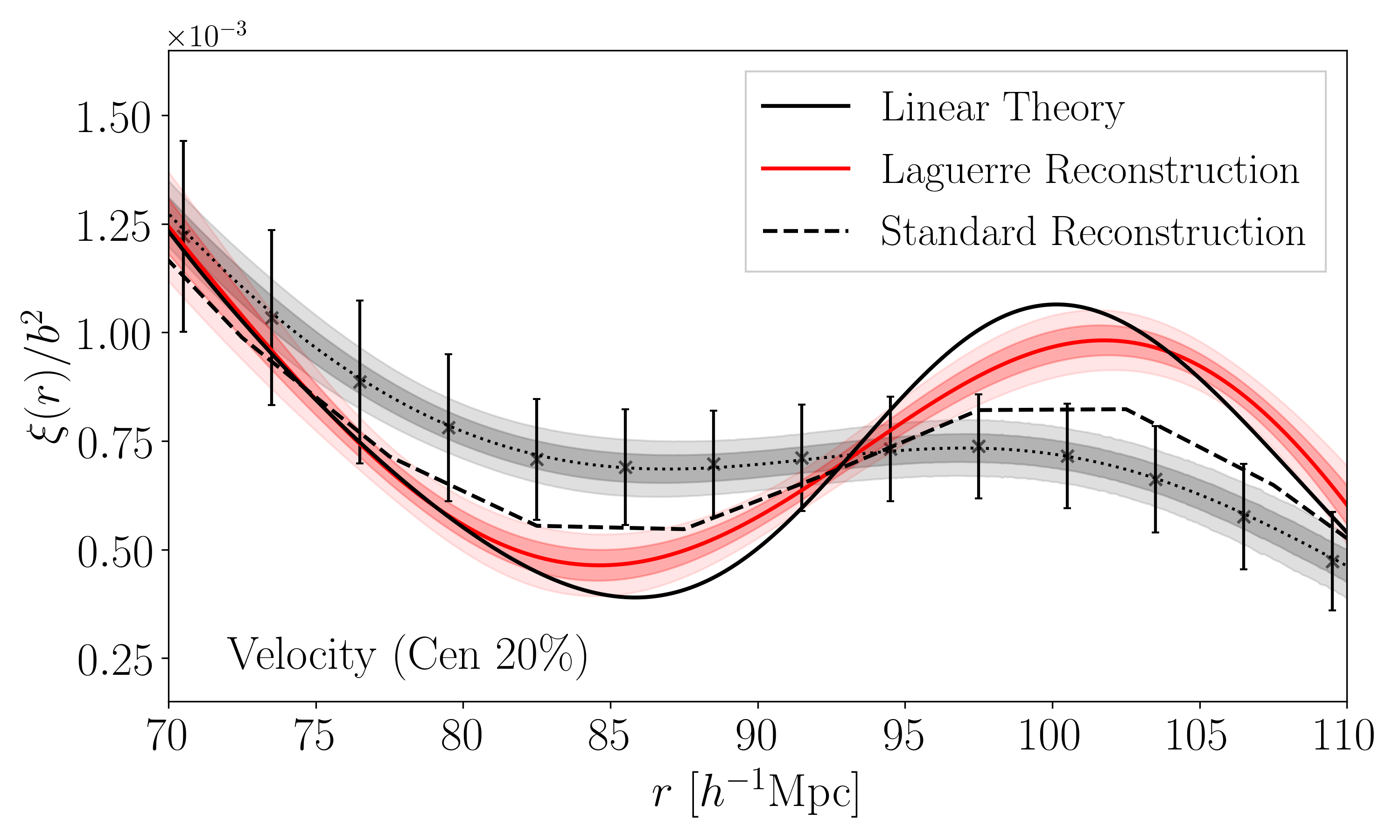}
 \includegraphics[width=0.45\hsize]{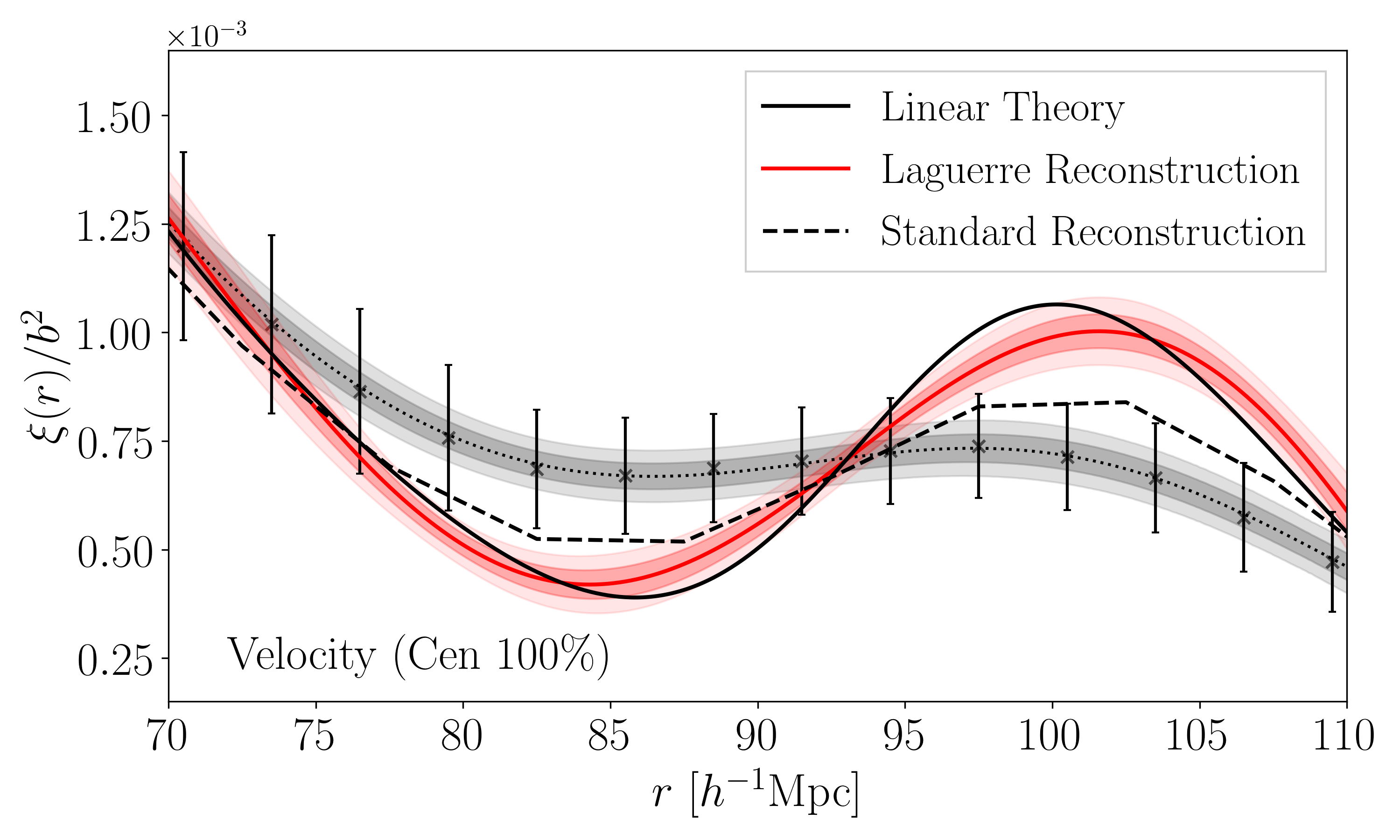}
 \caption{Comparison of the measured redshift-space monopole (symbols with error bars) in an effective volume of $20\times (1.1h^{-1}$Gpc)$^3$, with the 9th order Laguerre fit to it (grey bands show 68\% and 86\% confidence regions); the associated Laguerre reconstruction (pink bands); the linear theory correlation function (solid black); and the `standard' reconstruction from DE2019 (dashed).}
 \label{fig:shape}
\end{figure*}

\subsection{Idealized analysis:  Perfect prior information}\label{sec:ideal}
We present results for the larger effective volume first.
The symbols with error bars in Figure~\ref{fig:shape} show the average correlation functions for the four HODs.  We follow \cite{LPlaguerre} and fit these ($3h^{-1}$Mpc binned) correlation functions over the range 60-120$h^{-1}$Mpc to a set of half-integer Laguerre functions, up to 9th order.  The fitting makes use of the covariance matrix of the binned counts.  For the mock galaxy samples, there are 12 realizations of each HOD in each of the 20 simulation boxes, for a total of 240 correlation functions from which to estimate the (HOD-dependent) covariance matrix.  \rev{In practice, these covariance matrices are well approximated by the analytic estimate described in \cite{LPnus, LPlaguerre}, provided we use the tracer number density and bias factor that is appropriate for each HOD, and we set the survey volume equal to $20\times (1.1h^{-1}$Gpc)$^3$.}  Since the analytic covariance matrices are smoother, we use them when fitting.  The grey bands in the Figure~\ref{fig:shape} show the 68\% and 95\% confidence regions around the best fitting curves.  

We quantify the goodness-of-fit by computing $\chi^2$/dof.  Table~\ref{tab:results} summarizes the results:  it shows the effective bias factors of each sample, and the $\chi^2$/dof values, which indicate that quantities derived from these fits are likely to be meaningful.

\begin{figure}
    \centering
    \includegraphics[width=0.9\hsize]{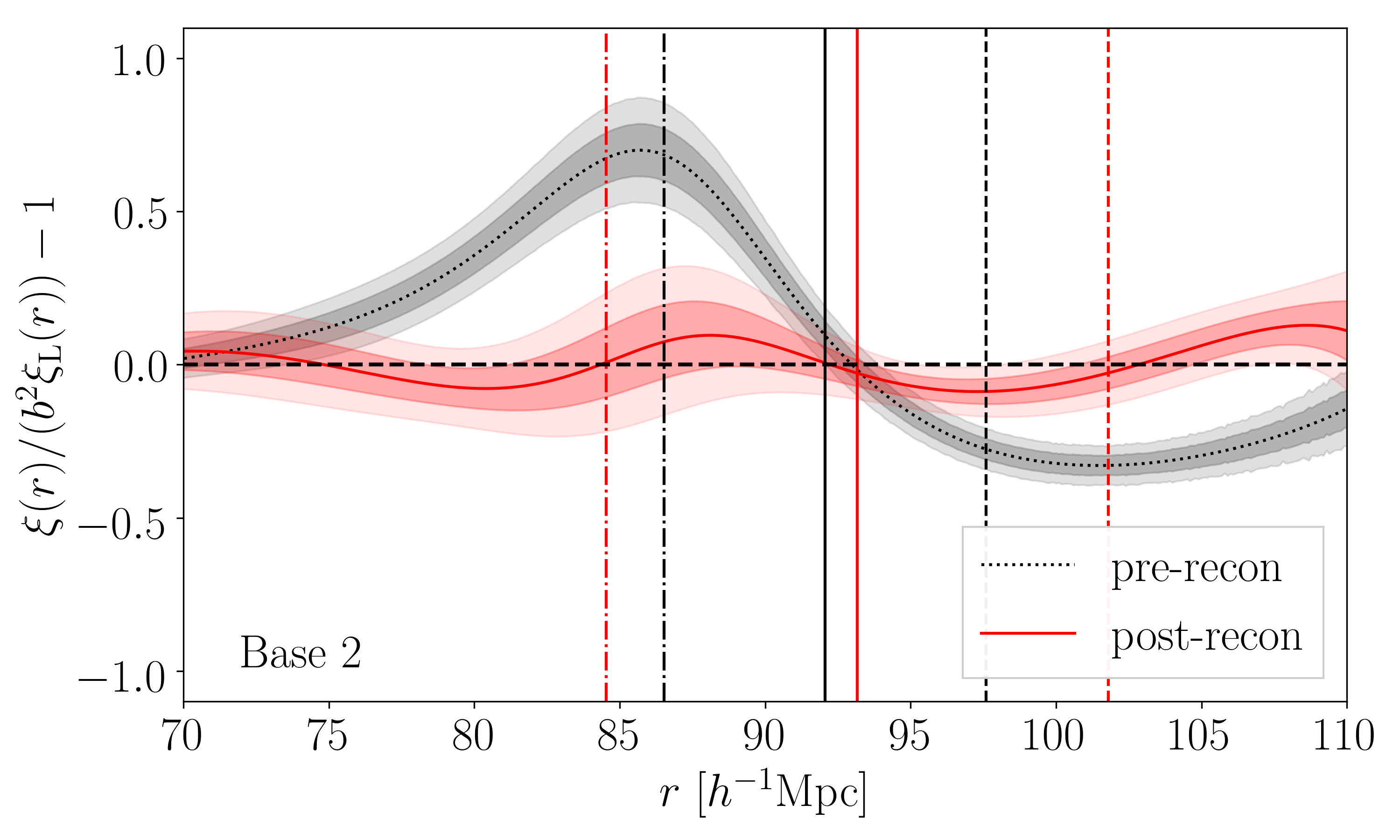}
  \caption{Ratio of the amplitudes of the measured $\xi$ (grey bands) and Laguerre reconstructed $\xi_{\rm Lag}$ (pink bands) to linearly biased $\xi_{\rm Lin}$ for the HOD model shown in the top left panel of Figure~\ref{fig:shape}.  Results for the other HODs are similar.  The amplitude of $\xi_{\rm Lag}$ is within 5\% of linear theory, except between $r_{\rm min}$ and $r_{\rm LP}$. \rev{Vertical lines respectively from left to right show the scales of $r_{\rm min}, r_{\rm max}$ and $r_{\rm LP}$ for pre- and post-reconstruction correlation functions.}}
  \label{fig:amp}
\end{figure}

The Laguerre reconstructed shape $\xi_{\rm Lag}$ results from choosing a value for the bias of the tracers and an estimate for the smearing scale and then using the fitted coefficients to construct a 9th order simple polynomial.  The red solid curves (and surrounding pink bands) in Figure~\ref{fig:shape} show this $\xi_{\rm Lag}$ if we use the correct values for these quantities.  We discuss how incorrect values impact the results shortly.  (The pink bands come from standard propagation of errors on the fitted coefficients, and account for the fact that these errors are correlated.)  Dashed curves show the shape returned by the `standard' reconstruction of DE2019 \rev{(error bars are similar to those of the measurements, so we have not shown them)}.  Evidently, $\xi_{\rm Lag}$ is substantially closer to linear theory (solid black).  While the peak and dip positions in $\xi_{\rm Lag}$ have clearly been over-corrected, \rev{$r_{\rm LP}$ is their average (c.f. equation~\ref{eq:rLP}), so it may still be accurate \cite[also see discussion in][]{PaperI}.}  

We check this explicitly by estimating $r_{\rm LP}$ of equation~(\ref{eq:rLP}) where $r_{\rm peak}$ and $r_{\rm dip}$ are the roots of $\partial\xi/\partial r=0$ -- with the derivative being computed analytically using the fitted coefficients.  Note that the same coefficients multiply different functions for the $\xi_{\rm Lag}$, so $r_{\rm LP}$ values pre- and post-Laguerre reconstruction can be different.  This is also true for the inflection point, $r_{\rm infl}$, which we estimate as the scale where $\partial^2\xi/\partial r^2 = 0$.  Table~\ref{tab:results} shows that, prior to reconstruction, $r_{\rm LP}$ and $r_{\rm inf}$ are offset from linear theory by about 1.5\%.  This shift is larger than it was for the halos in real space (shown in \cite{LPlaguerre}), in part because of the extra smearing that is due to redshift space distortions (equation~\ref{eq:sigeff}).  (Most previous LP analyses increase the measured LP by a factor of 1.005 \cite[e.g.][]{PaperI, PRDmocks, PRLboss, LPnus}.  We do not; but if we had, the result would still be offset from linear theory by about 1\%.)  Nevertheless, the Laguerre reconstructed values are substantially closer to linear theory indicating that it works well even in the presence of redshift-space distortions.  Moreover, the four HOD models agree to subpercent precision, both pre- and post-reconstruction.  Evidently, the LP is indeed rather independent of the biasing scheme (although it should be noted that the range of bias factors probed by the DE2019 HODs is small, so this is by no means an exhaustive test of the robustness of the LP).  

To emphasize the similarity between the Laguerre reconstructed correlation functions and linear theory, Figure~\ref{fig:amp} shows the ratio of the measured $\xi$ (grey bands), and the reconstructed $\xi_{\rm Lag}$ (pink bands) both normalized by $b^2\,\xi_{\rm Lin}$.  Vertical lines show the scales of $r_{\rm min}, r_{\rm max}$ and $r_{\rm LP}$ for pre- and post-reconstruction correlation functions.  This makes clear that, except between $r_{\rm min}$ and $r_{\rm LP}$, the reconstructed $\xi_{\rm Lag}$ is within about 5\% of $\xi_{\rm Lin}$.  Moreover, the error bands on the reconstruction show that $\xi_{\rm Lag}$ potentially provides better than 10\% constraints on the amplitude of $\xi_{\rm Lin}\propto (b\sigma_8)^2$.   

\begin{figure}
  \centering
  \includegraphics[width=0.9\hsize]{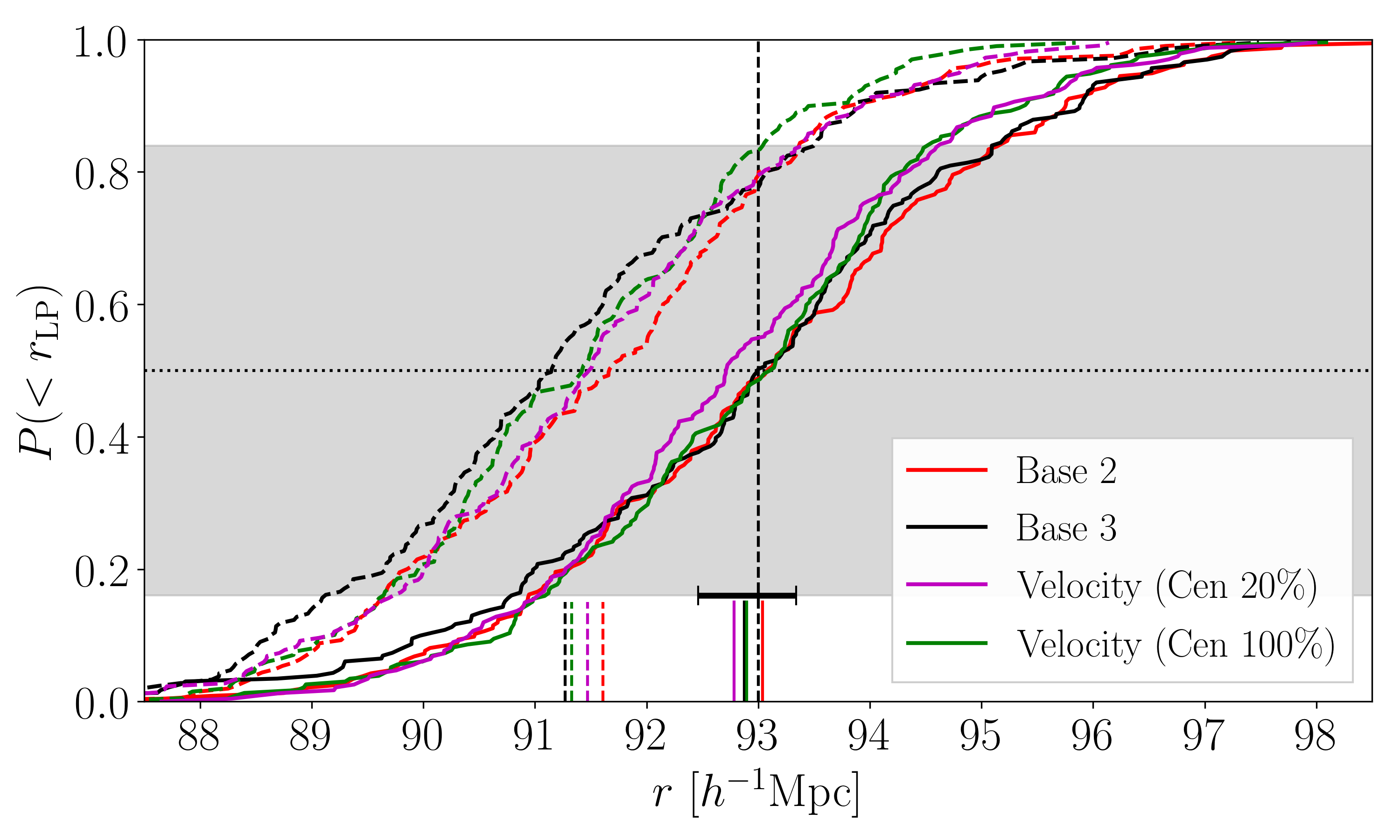}
  \caption{Distribution of LP scales estimated from the redshift-space monopole, pre- (dashed) and post- (solid) Laguerre reconstruction, for four different HOD models, in a $(1.1h^{-1}$Gpc)$^3$ volume.  The linear theory value is $93h^{-1}$Mpc.  Short vertical dashed lines show the four mean values, and horizontal bar shows the scale and its uncertainty reported in Table~\ref{tab:results} (i.e. determined from the correlation functions shown in Figure~\ref{fig:shape}, and corresponding to an effective volume that is $20\times$ larger than that of each simulation which contributes to this plot). Grey bands show the region which encloses 68\% of the values around the median.}
  \label{fig:cdf}
\end{figure}

We now present results from fitting to the simulations individually.  For a given HOD, the estimated $r_{\rm LP}$ values can differ from one another because of cosmic variance.  Figure~\ref{fig:cdf} compares the resulting distribution of $r_{\rm LP}$ values, pre- and post-reconstruction, in the 240 realizations for each HOD.  The four vertical bands bars at the bottom show the means of each distribution.  The horizontal vertical bars above them show the value of $r_{\rm LP}$ returned from fitting to the average of the correlation functions (reported in Table~\ref{tab:results}).  This error bar is about $4\times$ smaller than the 68\% range highlighted in the Figure, consistent with the fact that it was determined from an effective volume that was $20\times$ larger.  

\begin{figure}
  \centering
  \includegraphics[width=0.9\hsize]{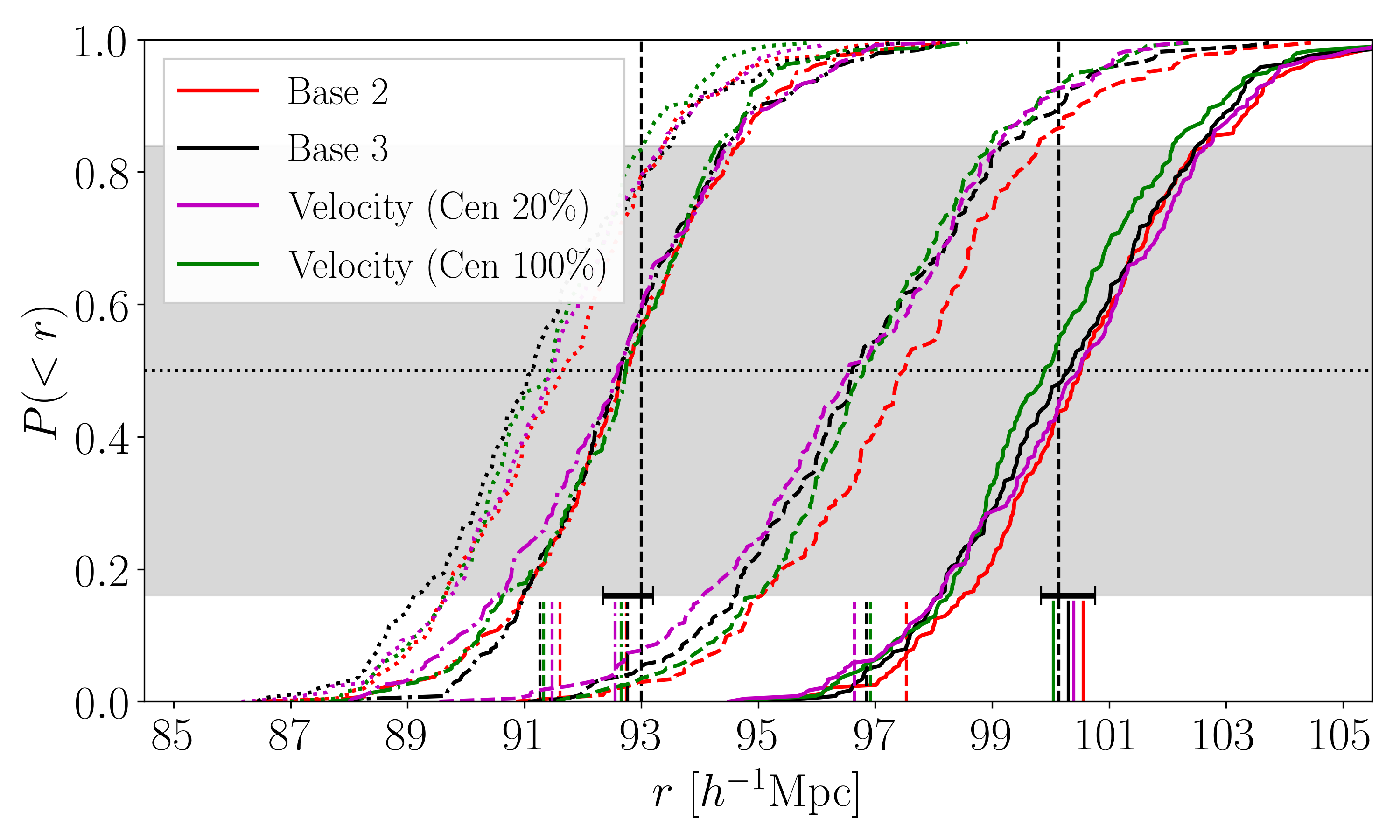}
  \caption{Similar to previous figure, but now for standard, density field reconstruction.   Dotted curves show the $r_{\rm LP}$-pre values (same as previous figure) and dot-dashed curves show the LP values estimated from the correlation functions in the `standard', density field reconstructions, even though the LP was not used to calibrate the reconstructions.  Dashed and solid curves show similar results for the peak scale ($r_{\rm peak} = 100h^{-1}$Mpc in linear theory) pre- and post-reconstruction.}
  \label{fig:cdfPeak}
\end{figure}


Figure~\ref{fig:cdfPeak} shows a similar comparison, but now for the scale $r_{\rm peak}$ which is close to the quantity which more traditional `density field reconstruction' algorithms -- in this case the `standard' reconstruction of DE2019 -- seek to reconstruct.  (The actual distance estimate is more sophisticated; it is based on fitting a template shape to a range of scales around $r_{\rm peak}$, \rev{and is the subject of Appendix~\ref{sec:alphas}}.)  The solid curves show $r_{\rm peak}$ measured in the `standard' reconstruction of $\xi$ (i.e. we fit half-integer Laguerre functions to the dashed curves in Figure~\ref{fig:shape}, and estimate $r_{\rm peak}$ from the fit).  This shows that $r_{\rm peak}$ is indeed accurately reconstructed, with approximately the same precision that Laguerre reconstruction achieves for $r_{\rm LP}$ (the $x$-axis here covers about $20h^{-1}$Mpc, whereas the previous figure covered about $10h^{-1}$Mpc). \rev{Despite the fact that this `standard' reconstruction (dot-dashed curves) does not reconstruct the shape of $\xi$ over a wide range of scales, it does do a good job of reconstructing $r_{\rm LP}$ -- with a small bias that is slightly smaller than the error bar -- even though this is not something it was calibrated to do. }


These plots illustrate nicely the virtues of working with the LP:  prior to reconstruction, it is shifted from linear theory by substantially less than $r_{\rm peak}$ \rev{(median $r_{\rm LP}$ offset is $\sim 1.5h^{-1}$Mpc vs $3h^{-1}$Mpc) for $r_{\rm peak}$}, and the variations between HODs are slightly smaller \rev{(median values of the HODs are within $0.5h^{-1}$Mpc of one another for LP, but the scatter is twice as large for $r_{\rm peak}$)}.  However, the main point of this comparison is to show that, despite its simplicity, Laguerre reconstruction of the two-point correlation function enables distance scale estimates that are comparable in precision and accuracy to density field reconstruction methods, \rev{in the ideal case in which both methods assume perfect prior information}.

\subsection{Impact of uncertain prior information}\label{sec:realistic}

\begin{figure}
 \centering
 \includegraphics[width=0.45\hsize]{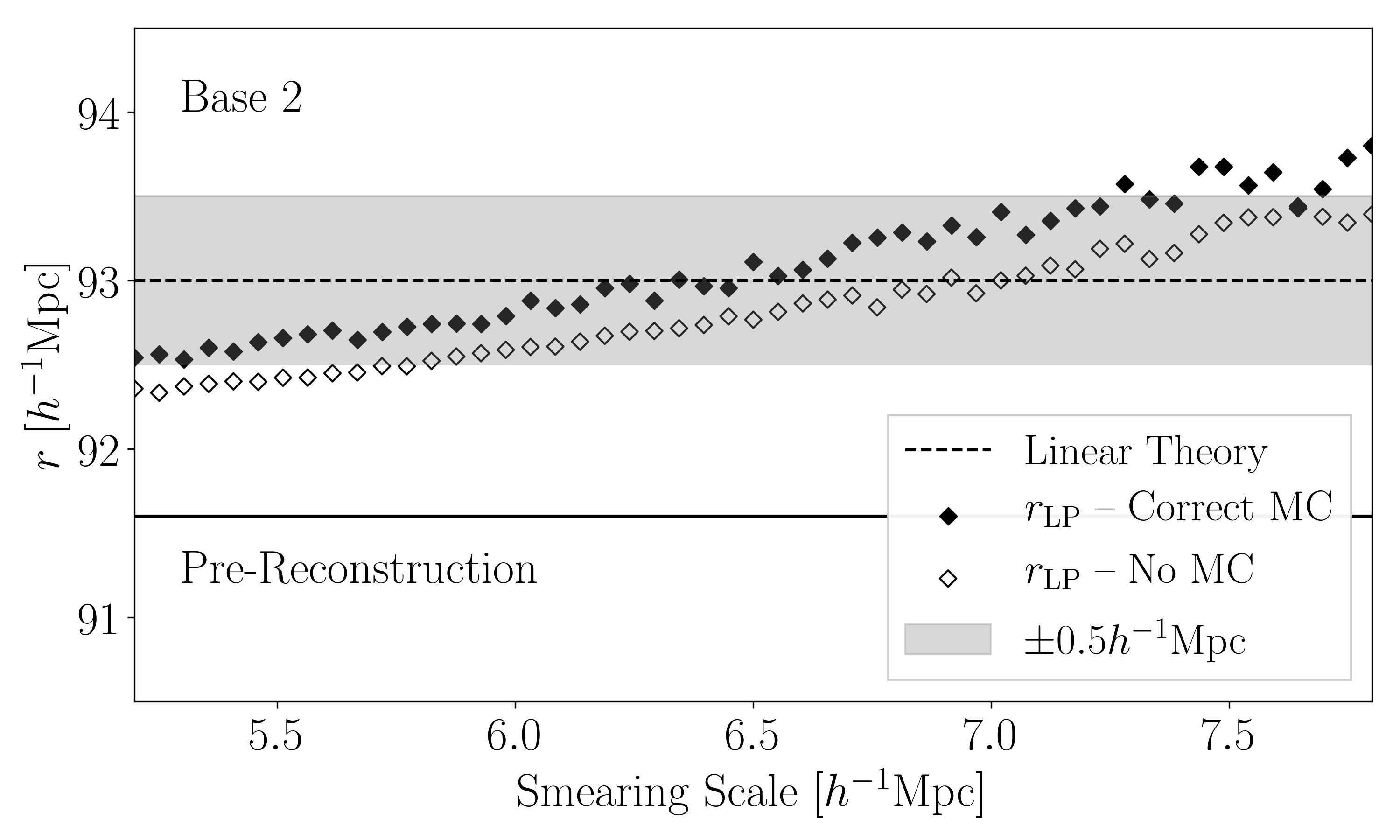}
 \includegraphics[width=0.45\hsize]{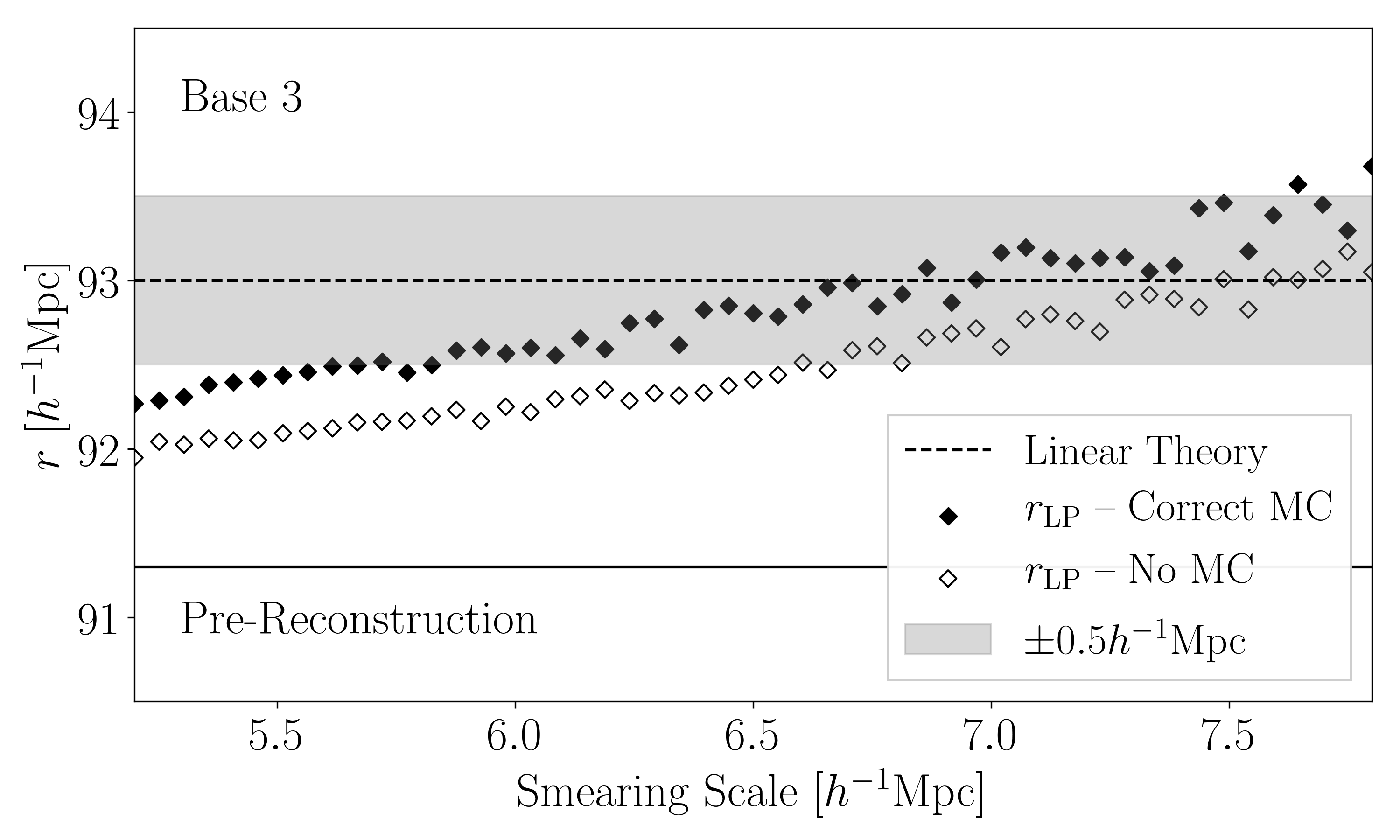}
 \includegraphics[width=0.45\hsize]{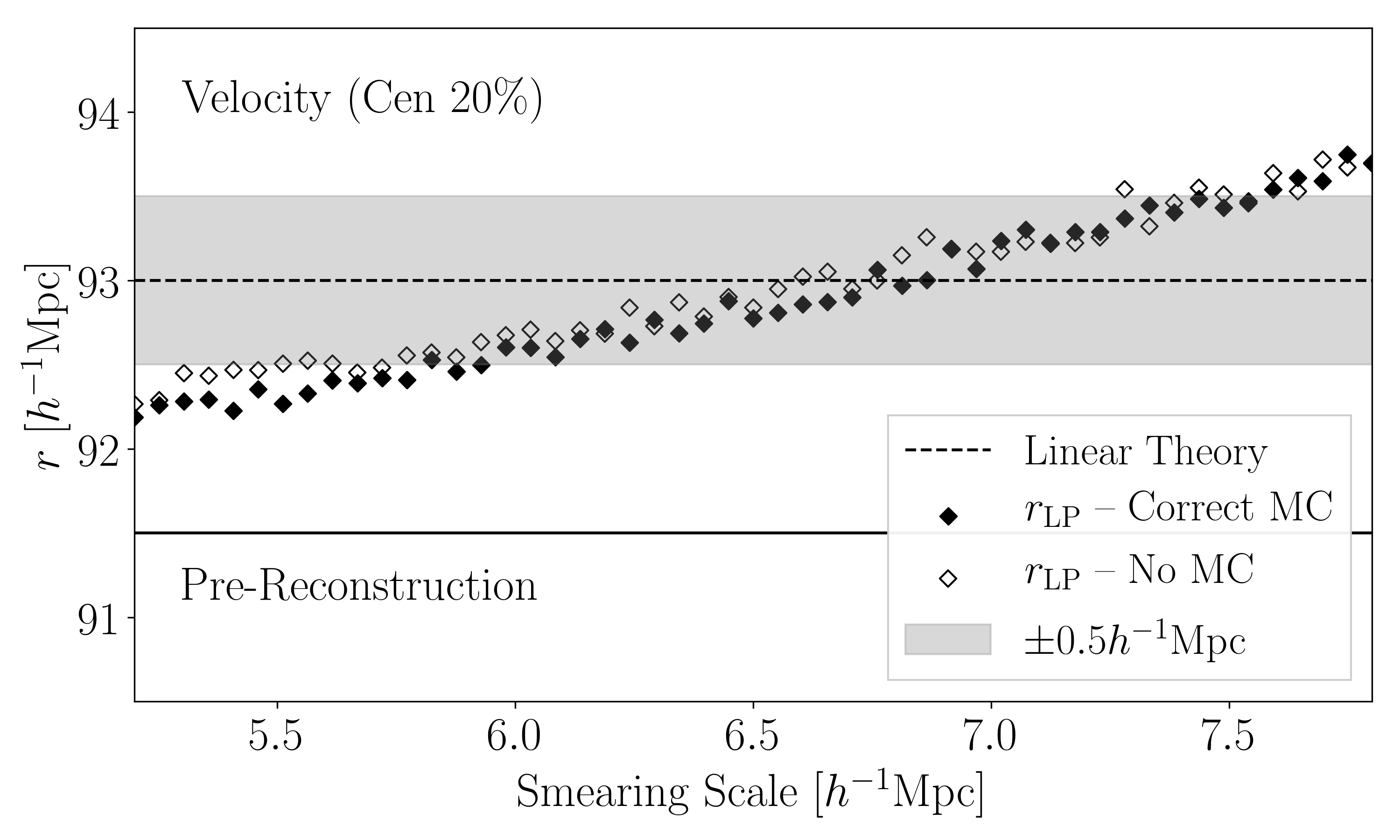}
 \includegraphics[width=0.45\hsize]{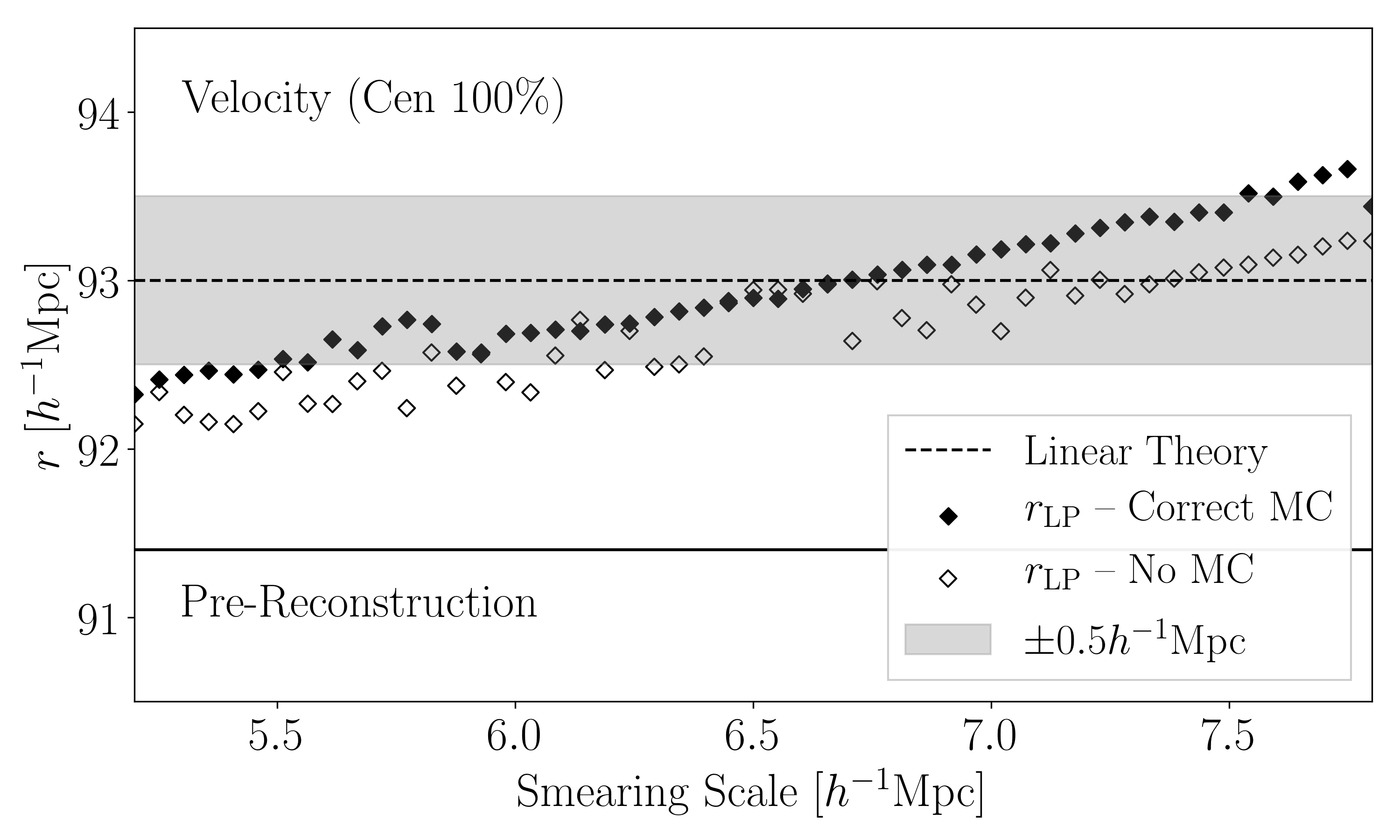}
 \caption{Dependence of $r_{\rm LP}$ in $\xi_{\rm Lag}$ on assumed smearing scale.
   Unfilled symbols do not account for mode-coupling, so require no information about galaxy bias; filled symbols use the correct value of galaxy bias when accounting for mode-coupling.}   
 \label{fig:noMC}
\end{figure}

Laguerre reconstruction depends on an assumed bias factor and smearing scale.  We now explore how the reconstructed $r_{\rm LP}$ scale is affected if we use incorrect values, as is likely to occur in real data.  (Density field reconstruction algorithms must make analogous choices.)  We do this in two steps because, in fact, the bias factor is only necessary if one wishes to account for what is known as mode-coupling \cite{LPlaguerre}. Therefore, Figure~\ref{fig:noMC} studies the impact of this term on the reconstruction.  Filled symbols show the dependence on the smearing scale when this term is accounted for using the correct bias factor, and open symbols show the result of neglecting this term completely.  There is a difference, but it is not large.  This is encouraging since it suggests that the cost to the method of being completely agnostic about the value of $b$ is not prohibitive.  

\begin{figure}[b]
 \centering
 \includegraphics[width=0.47\hsize]{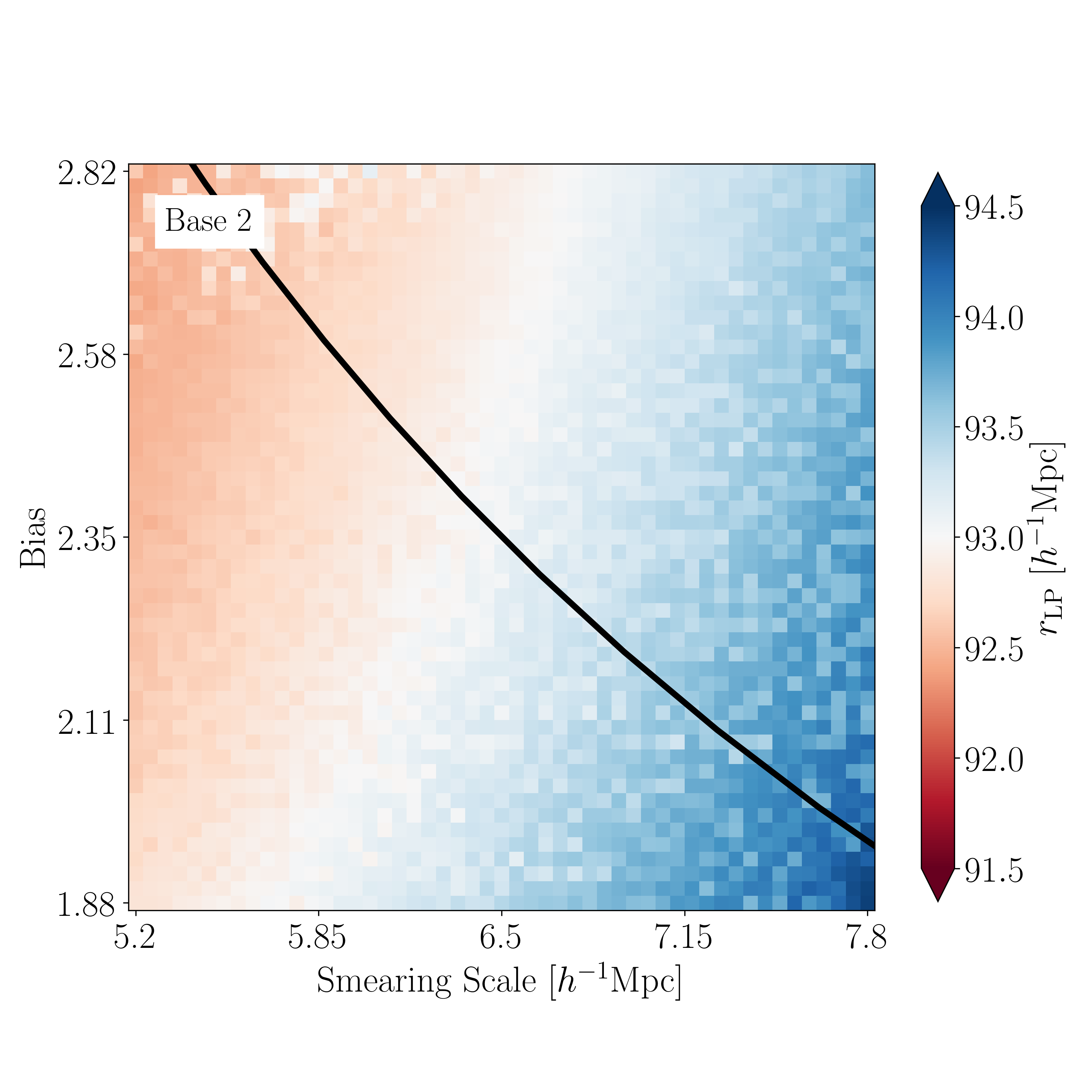} 
 \includegraphics[width=0.47\hsize]{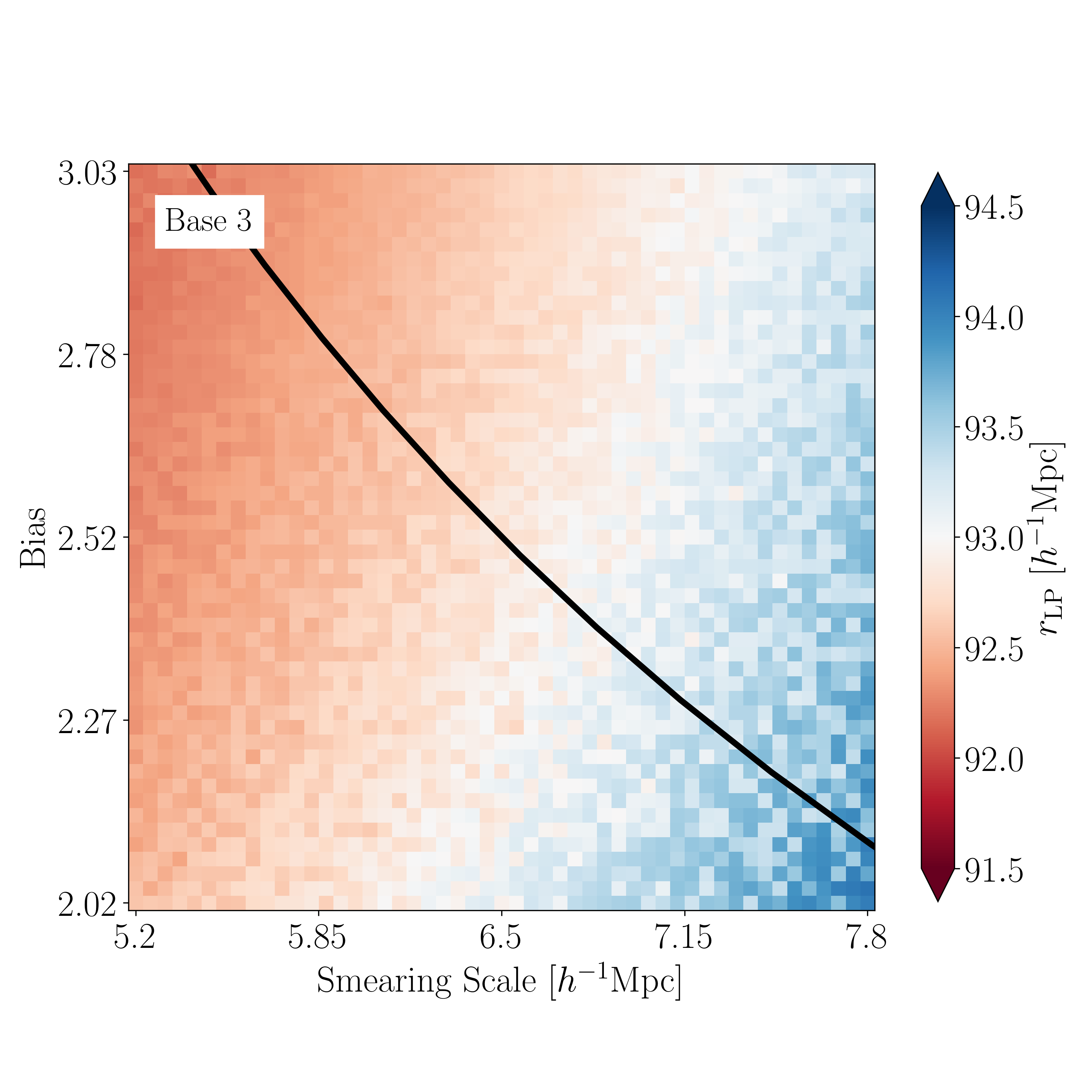} 
 \includegraphics[width=0.47\hsize]{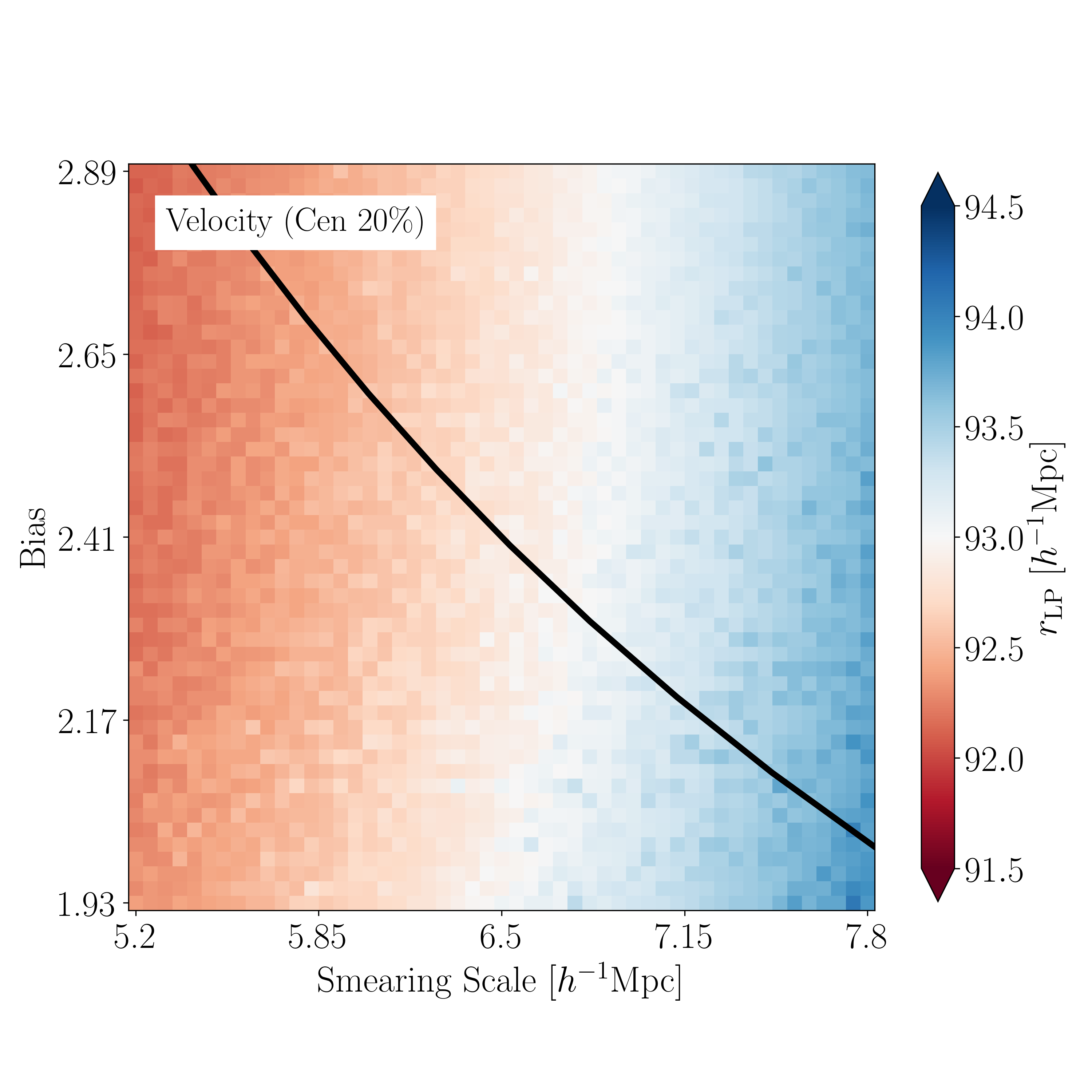}
 \includegraphics[width=0.47\hsize]{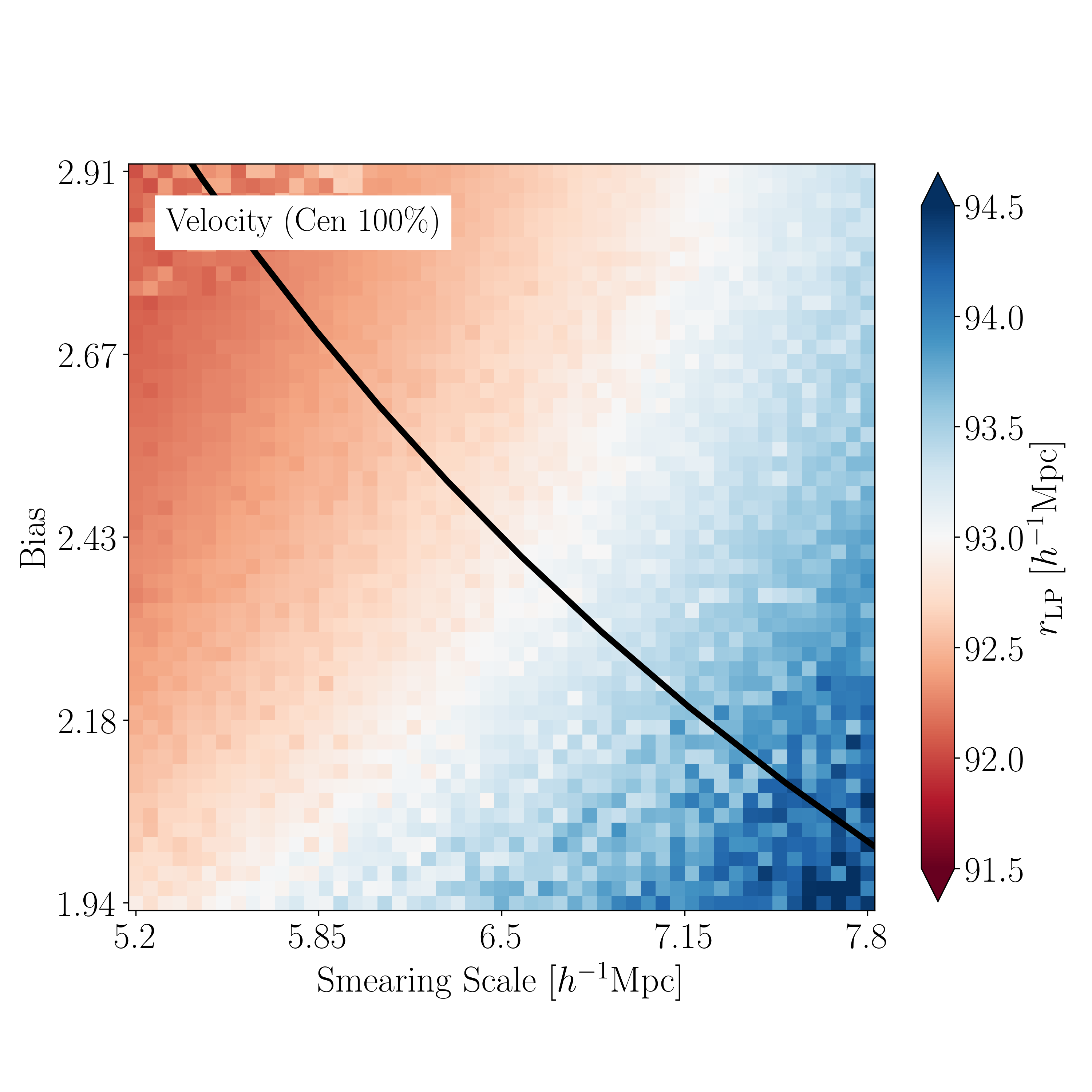}
 \caption{Dependence of $r_{\rm LP}$ measured in $\xi_{\rm Lag}$ on assumed bias factor and smearing scale.  \rev{White pixels show the combinations which return unbiased reconstructions of the linear theory value ($93h^{-1}{\rm Mpc}$).}
   Black curves show $b\propto \Sigma^{-1}$, centered on the correct values of $b_{\rm eff}$ and $\Sigma_{\rm eff}$ for each panel.  
In $\Lambda$CDM models, this is the locus along which one should read off $r_{\rm LP}$ values so as to get more realistic uncertainties on $r_{\rm LP}$.  If $b_{\rm eff}$ or $\Sigma_{\rm eff}$ are unknown then this will shift the curves to the left or right, potentially broadening the error estimate further.}
 \label{fig:ba}
\end{figure}

Figure~\ref{fig:ba} shows how the estimated LP scale depends on the assumed bias and smearing scale.  The panels for each HOD show 20\% variations in each direction around the correct value, both for $b$ and for $\Sigma$; realistic uncertainties are likely to be smaller.  All panels show a degeneracy in the sense that increasing the smearing scale and the bias leave the inferred LP unchanged.

Figure~\ref{fig:ba} indicates that constraints on the distance scale from a survey with comoving volume $27 h^{-3}$Gpc$^3$ will not be as tight as Table~\ref{tab:results} suggests.  Rather, realistic constraints on the distance scale will require averaging over plots like these, weighting by priors on the values of $b$ and $\Sigma$.  Note that these are unlikely to be circular averages in the $b-\Sigma$ plane because the priors are likely to be correlated.  Rather, realistic uncertainties on $r_{\rm LP}$ are likely to be associated with averaging along $b\propto \Sigma^{-1}$ \cite{LPlaguerre}.  The thick black curves in Figure~\ref{fig:ba} show this scaling, centered on the correct value of $b_{\rm eff}$ and $\Sigma_{\rm eff}$ for each panel.  

\begin{figure}[b]
  \includegraphics[width=0.47\hsize]{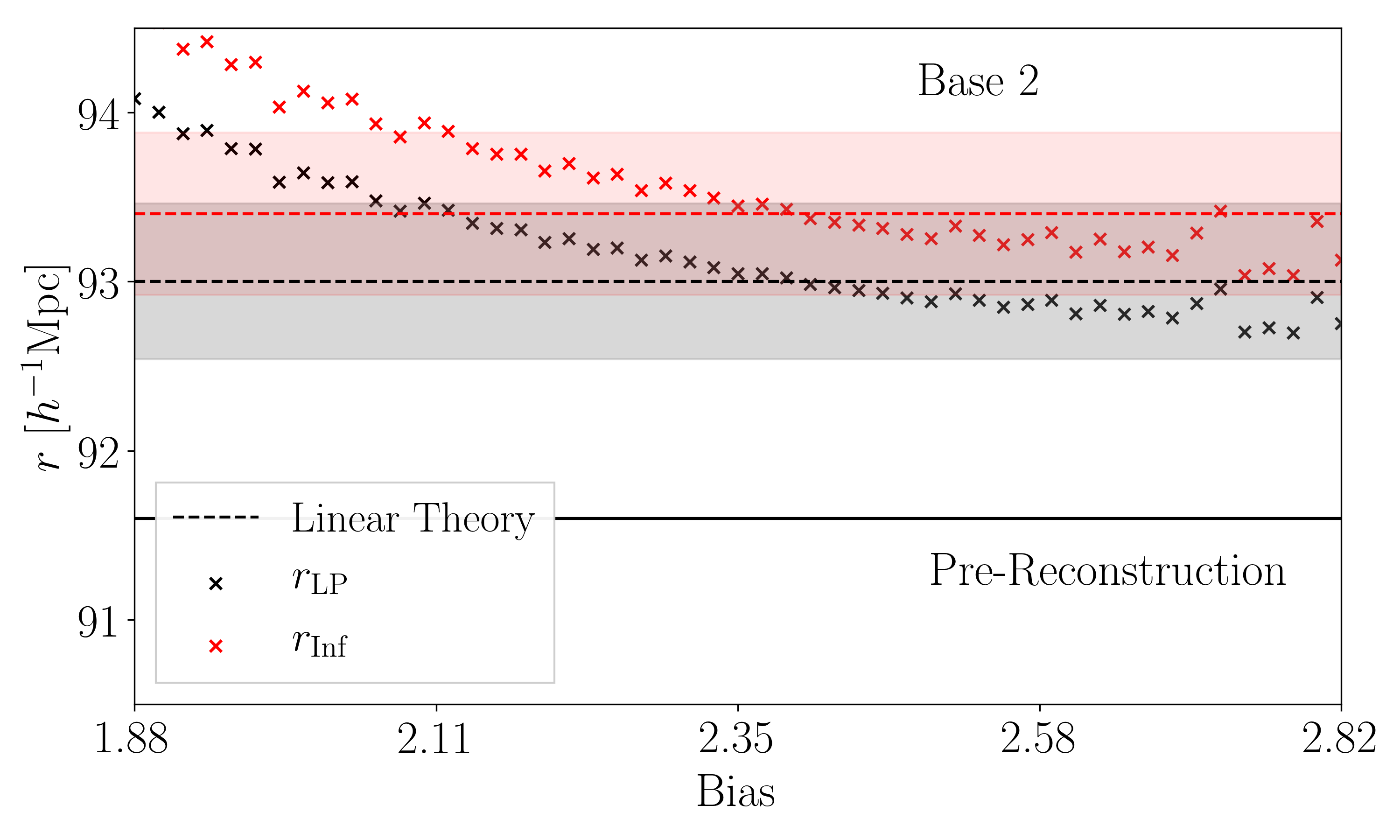}
  \includegraphics[width=0.47\hsize]{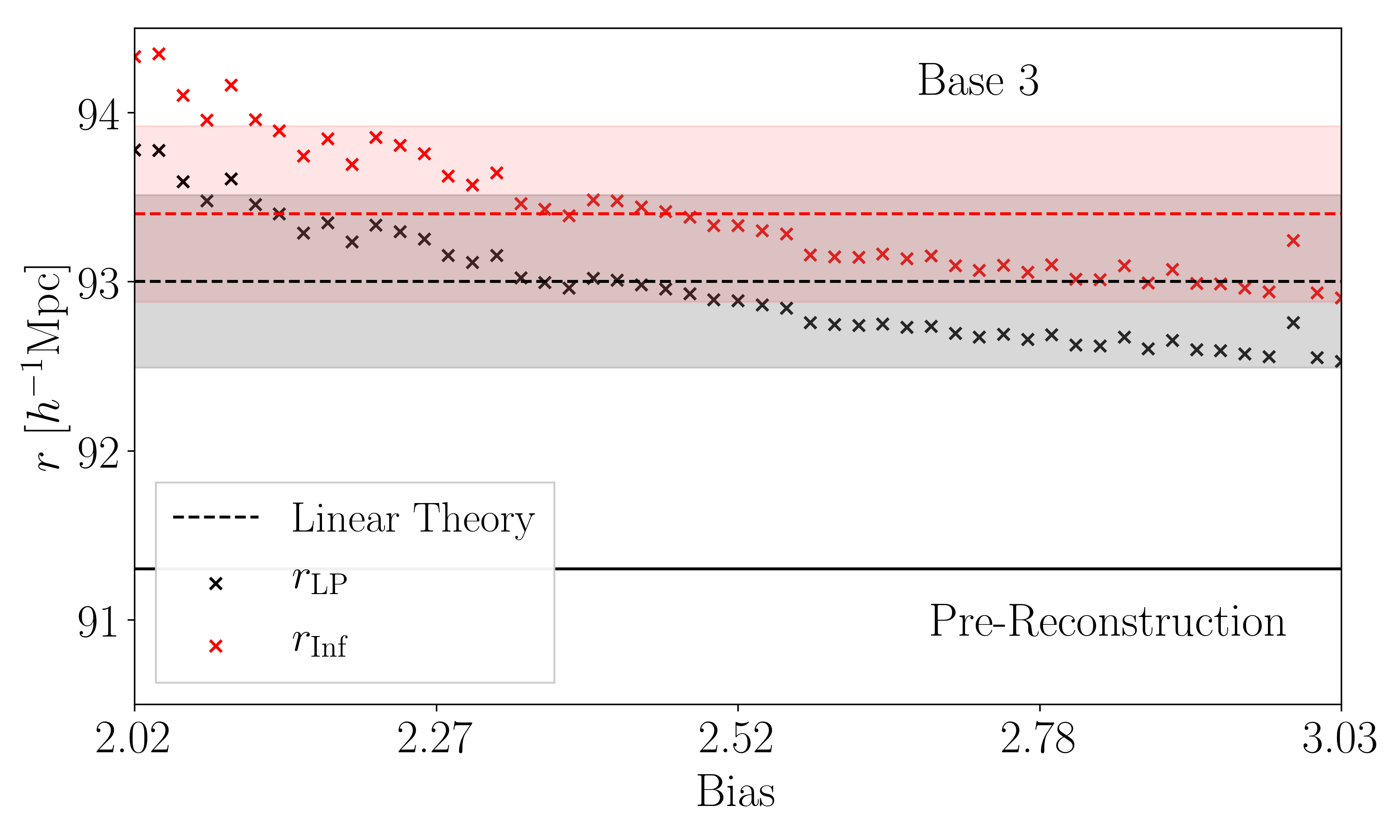}
  \includegraphics[width=0.47\hsize]{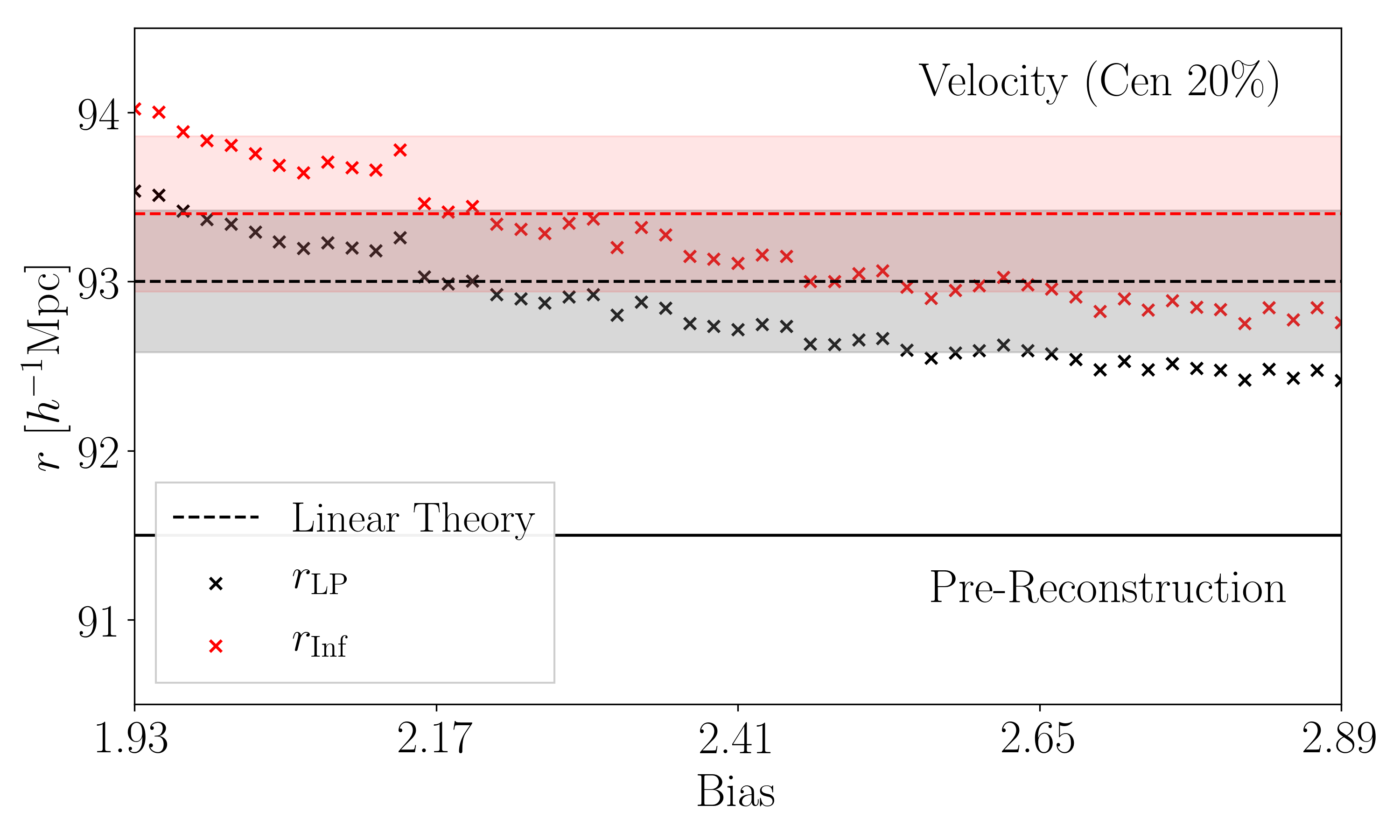}
  \includegraphics[width=0.47\hsize]{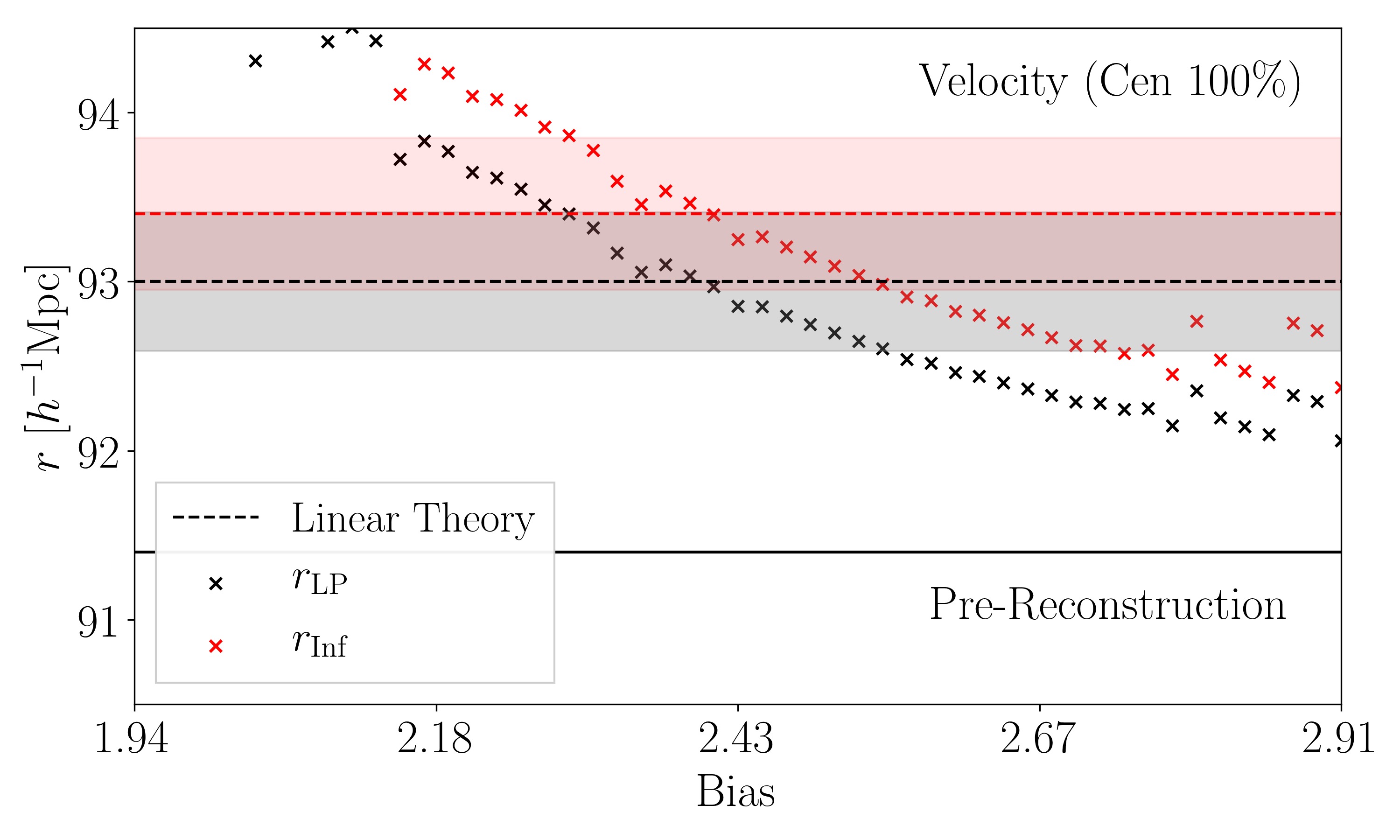}
  \caption{Inferred $r_{\rm LP}$ and $r_{\rm infl}$ scales as one moves along the black curves shown in the previous figure.  Pink and grey bands show the uncertainties quoted in Table~\ref{tab:results}, which assume that $b_{\rm eff}$ and $\Sigma_{\rm eff}$ are known perfectly:  accounting for the fact that they are not broadens the uncertainty on the distance scale.}
  \label{fig:rLPb}
\end{figure}

Figure~\ref{fig:rLPb} shows how $r_{\rm LP}$ and $r_{\rm infl}$ vary as one averages along the black curves shown in Fig.~\ref{fig:ba}.  The colored bands show the error bars in Table~\ref{tab:results} which assume $b_{\rm eff}$ (and $\Sigma_{\rm eff}$) are known perfectly.  The fact that some symbols lie outside these bands shows that accounting for uncertainties in these parameters can broaden the errors on the inferred distance scale if the allowed range in $b_{\rm eff}$ is sufficiently wide. This is particularly true for the HOD shown in the bottom right panel (but recall that this is a rather extreme case: \rev{presumably, the larger smearing in this model requires more reconstruction, making the choices of $b_{\rm eff}$ and $\Sigma_{\rm eff}$ more critical}).  Allowing for the additional (survey-specific) uncertainty in the overall normalization of the black curves will further degrade the constraints.  However, because lines of fixed $r_{\rm LP}$ run approximately perpendicular to the black curves in each panel, the degradation in constraining power may not be prohibitive.

\subsection{Synergy between linear point and density field reconstruction}\label{sec:synergy}
The previous subsection made the point that realistic errors on the distance scale require marginalization over the parameters which are input to Laguerre reconstruction.  For similar reasons, constraints from density field reconstruction algorithms which assume a single set of parameters (e.g., background cosmology, $\sigma_8$, bias), likely underestimate the true uncertainties, if they do not include the effect of marginalizing over the uncertainties in these input parameters \cite{PRDforecast}.  Since none of the error bars reported in Table~\ref{tab:results} (nor their counterparts in DE2019) include such marginalization, they likely underestimate the true uncertainties on the distance scale.  

The importance of this additional marginalization for density field reconstruction has yet to be completely quantified, in part because doing so is computationally demanding \rev{\cite[see][for important first steps towards this goal]{vm+2018}}.  This is because such methods have two steps:  reconstruction, followed by estimation of the distance scale from the reconstructed field.  Typically this second step uses a template model which is fit to the two-point statistics of the reconstructed field.  In principle, as one marginalizes over the parameters used to perform the reconstruction, one must take care to self-consistently modify the template as well.  However, Fig.~\ref{fig:cdfPeak} shows that the Linear Point -- which can be estimated without this extra `self-consistent template' step -- provides a simpler route to the distance scale in reconstructed fields.  In particular, because the Linear Point is rather insensitive to one of the dominant sources of uncertainty -- galaxy bias -- our results suggest that the Linear Point may be useful in determining more realistic error estimates for density field reconstruction methods.  In this respect, the utility of the Linear Point transcends that of Laguerre reconstruction.  

\section{Summary}
We have shown that Laguerre reconstruction of the monopole of the redshift-space distorted correlation function substantially mitigates the effects of nonlinear evolution for a variety of interesting halo-based biasing schemes (Figures~\ref{fig:shape} and~\ref{fig:amp}).  The Linear Point (equation~\ref{eq:rLP}) in the reconstructed correlation function provides comparable accuracy and precision to the distance scale estimates provided by `standard' density field reconstruction algorithms (Figures~\ref{fig:cdf} and~\ref{fig:lag+std}, and Table~\ref{tab:results}).

Like all other reconstruction schemes, Laguerre reconstruction depends on certain input parameters (a smearing scale and a galaxy bias factor).  If these are not known precisely, then realistic constraints on the distance scale will be broadened.  For Laguerre reconstruction, this degradation in constraining power is straightforward -- both conceptually and computationally -- to estimate (Figure~\ref{fig:ba} and associated discussion).  Performing the analogous marginalization over poorly constrained input parameters is more computationally demanding for density field reconstruction methods.  For these reasons, we believe Laguerre reconstruction in combination with the LP offers a simple and practical complementary estimate of cosmological distance scales.  In addition, measuring the LP in density field reconstructions simplifies the process of providing realistic error estimates on the distance scale from these methods (Section~\ref{sec:synergy}).  Therefore, our results show that the utility of the LP is not confined to Laguerre reconstruction.


\begin{acknowledgments} 
  We are very grateful to Y. T. Duan and D. Eisenstein for providing their measurements in a convenient electronic format, \rev{and the code they used for estimating the distance scale from these measurements}, and to N. Padmanabhan, G. Starkman and S. Anselmi for helpful discussions.  
  FN and RKS thank the Munich Institute for Astro- and Particle Physics (MIAPP) which is funded by the Deutsche Forschungsgemeinschaft (DFG, German Research Foundation) under Germany's Excellence Strategy – EXC-2094 – 390783311, for its hospitality during the summer of 2019.
  FN acknowledges support from the National Science Foundation Graduate Research Fellowship (NSF GRFP) under Grant No. DGE-1845298.
 IZ is supported by NSF grant AST-1612085.
\end{acknowledgments}

\bibliography{laguerreHOD_rev.bib}

\appendix

\section{Abacus and Emulator simulation sets}
\begin{figure}
 \centering
 \vspace{1cm}
 \includegraphics[width=0.9\hsize]{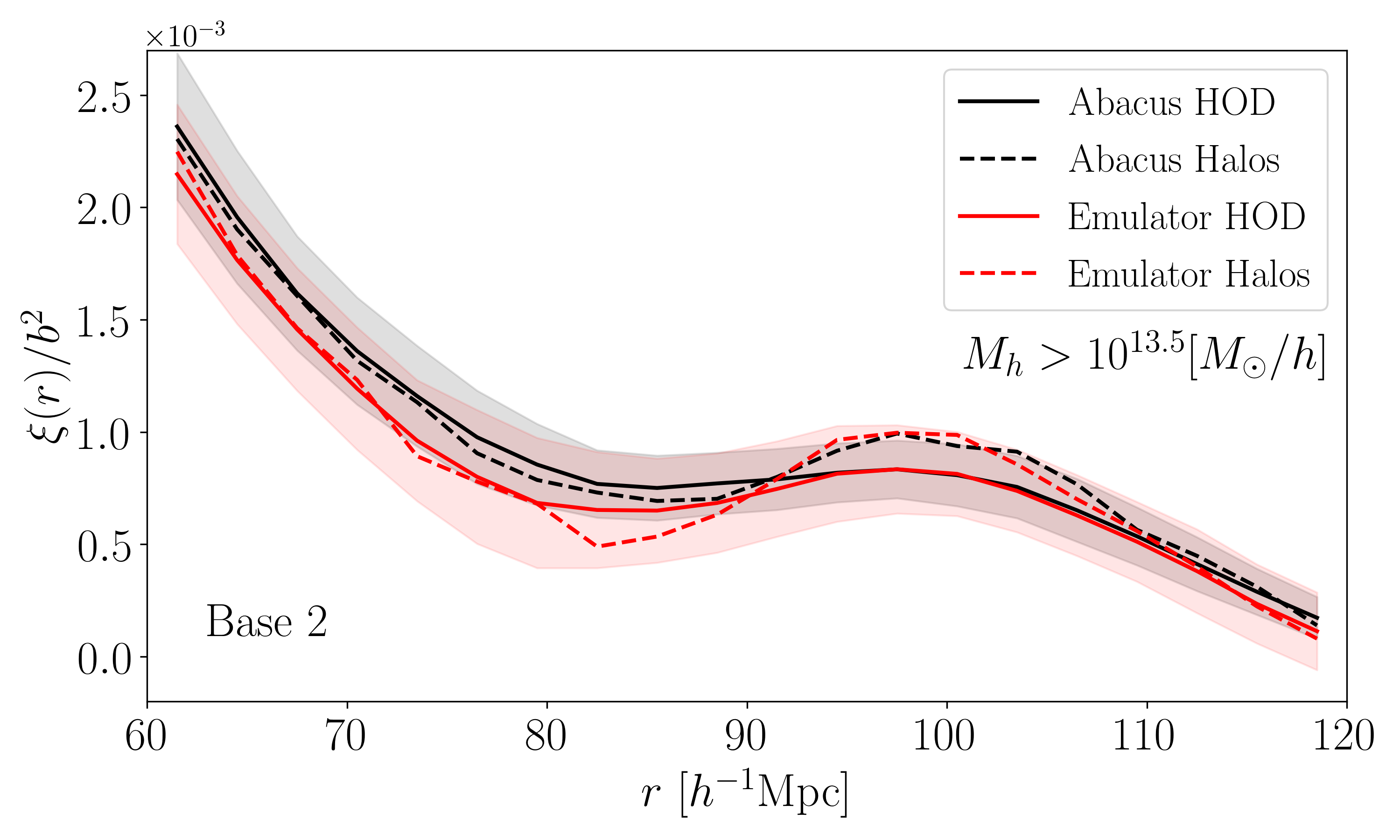}
 \caption{Real space correlation functions of massive halos in the Abacus and Emulator simulation sets (dashed curves) differ slightly around the dip scale.  The monopole of the redshift-space correlation function in the `base 2' HOD model shows similar differences (solid curves, bands show the 1$\sigma$ error bars). These differences are similar for all the HOD models considered in the main text. }
 \label{fig:AE}
\end{figure}

We did not use the additional 16 `Emulator' boxes that have the same background cosmology and are also part of the {\tt ABACUS COSMOS} release for reasons discussed in \cite{LPlaguerre}.  In particular, the two dashed curves in Fig.~\ref{fig:AE} compare the real-space correlation functions of the massive halo samples used by \cite{LPlaguerre} in the Abacus and Emulator simulation sets (effective volumes of $27h^{-3}$Gpc$^3$ and $21h^{-3}$Gpc$^3$, respectively).  While the agreement is good around the peak scale, the differences around the dip are larger, and only marginally consistent with cosmic variance.  These differences are also present in the redshift-space distorted HOD samples.  To illustrate, the two solid curves show the redshift space monopole of the same HOD population in the two simulation sets (we have normalized each halo sample to have the same amplitude as the associated HODs at $\sim 70h^{-1}$Mpc).  Comparison of the HOD and high mass samples (solid and dashed curves) also graphically illustrates the additional smearing of the BAO signal that is associated with the redshift space distorted signal, whose impact on Laguerre reconstruction we would like to assess.

\section{Full-shape constraints with perfect prior information}\label{sec:alphas}
The main text used the Linear Point to quantify the precision and accuracy on the inferred distance scale in the Laguerre-reconstructed correlation functions, and in those measured in the full density-field reconstructions.  The virtue of the LP is that it returns a distance scale estimate which does not require prior knowledge of the shape of the correlation function.  All other published BAO distance scale estimates rely on fitting a model template to the reconstructed correlation functions.  This clearly requires considerably more prior information, so it is natural to ask if fitting to the full shape returns considerably greater accuracy.

We address this in three steps:
First, we compare constraints from the LP with those from fitting to the full shape of the Laguerre-reconstructed redshift space monopole.  Next, we compare the $\alpha_0$ values in $\xi_{\rm Lag}$ with those in $\xi_0$ of the standard reconstruction.  Finally, we compare the constraints from the monopole $\xi_0$ with those from the full shape of the redshift-space correlation function.  This final step is only possible for the standard reconstruction:  if the gains in precision and accuracy are considerable, then this motivates extending both the Laguerre reconstruction and the Linear Point methodology to account for anisotropic redshift space distortions.

We begin by defining 
\begin{equation}
  \alpha_{\rm LP}\equiv \frac{93h^{-1}{\rm Mpc}}{r_{\rm LP}}
\end{equation}
and $\alpha_0$, which is the value for which 
\begin{equation}
  \xi_{\rm T}(\alpha_0 s) = B^2 \int \frac{dk}{k}\frac{k^3 P_{\rm T}(k)}{2\pi^2}
  \, j_0(ks\alpha_0) + ...
 \label{eq:xiTemp}
\end{equation}
(where the `...' refer to nuisance parameters which scale as $s^{-1}$ or $s^{-2}$; see DE2019 for details) best fits each of the measured $\xi_0(s)$ curves over the range $60-120h^{-1}$Mpc.  Here $P_{\rm T}(k)$ is the linear theory dark matter correlation from CAMB for the {\tt ABACUS} cosmology, and $B$ is a free parameter.  Note that $\alpha > 1$ means the estimated scale is smaller than the true one.

For fitting to the full shape, our goal is to parallel the analysis in DE2019 as closely as possible. Therefore, for their standard density field reconstructions, we use the same code and covariance matrices they used when fitting, which they kindly provided to us.  Their analysis first averages all the HODs in a single box, and then performs a jackknife resampling of the boxes to obtain smoother covariance matrices, which they then rescale to represent the volume of a single box.  Therefore, for our Laguerre analysis, we construct covariance matrices by first co-adding all 12 reconstructed $\xi_{\rm Lag}$ in a box, and we then average together 19 randomly chosen boxes at a time, repeating 50 times.  
The resulting covariance matrix is smooth, and we rescale it, as they do, to the volume of a single box, before using their fitting code to determine $\alpha_0$.  Note that for every bias model and every type of correlation there is a different covariance matrix.  However, we have checked that the results which follow are not changed substantially if we use the same covariance matrix for all the HODs (essentially because the HODs are not very different from one another).

\begin{figure}
 \centering
 \includegraphics[width=0.9\hsize]{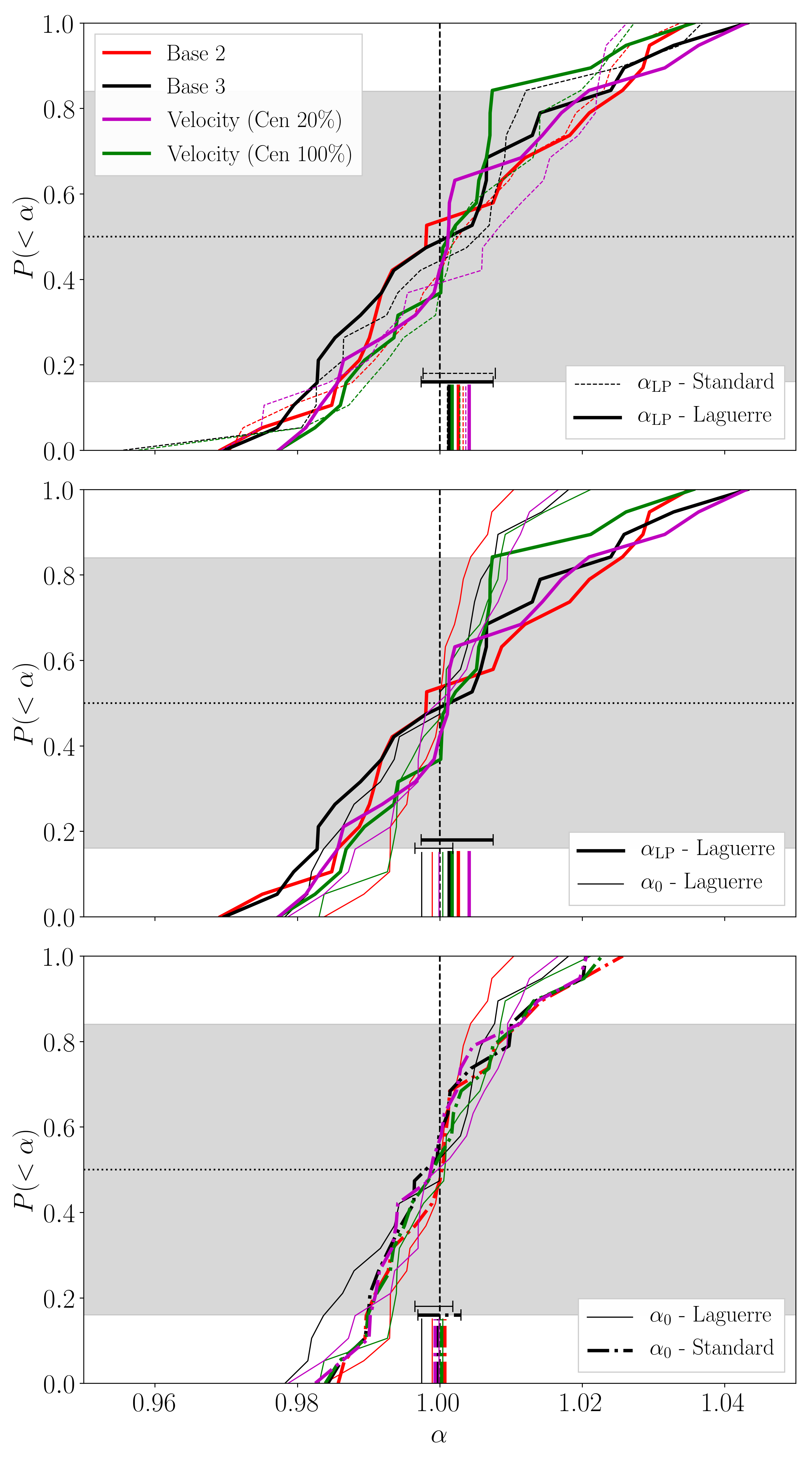}
 \caption{\rev{Top:  Cumulative distributions of $\alpha_{\rm LP}$ in the Laguerre (thick solid) and standard (thin dashed) reconstructions are very similar, indicating that the two reconstructions provide similar constraints on the distance scale. Middle: Cumulative distributions of $\alpha_{\rm LP}$ (thick solid, same as top panel) and $\alpha_0$ (thin solid) measured from $\xi_{\rm Lag}$. The distribution of $\alpha_0$ is narrower, indicating that fitting to the full shape of the redshift space monopole $\xi_0$ yields greater precision than the LP scale itself. Bottom: Cumulative distributions of $\alpha_0$ from fitting to $\xi_{\rm Lag}$ (thin solid, same as middle panel) or to $\xi_0$ in the standard reconstruction (dot-dashed). Consistent with the top panel, the two reconstructions provide similar constraints on the distance scale. Grey bands show the region which encloses 68\% of the values around the median. Bars along the bottom of each panel show the uncertainty associated with an effective volume which is 20 $\times$ larger.}}
 \label{fig:lag+std}
\end{figure}

We present our results in a format which is similar to Figure~\ref{fig:cdf} in the main text. However, in contrast to the main text, for which each HOD realization contributes separately (for a total of 240 $\alpha_{\rm LP}$ values for each HOD), here all 12 realizations of each HOD in a box are first coadded, so that we have only 20 values for each HOD (this is to mimic Fig. 2 in DE2019, which shows 36 points per HOD, from the 20 ABACUS + 16 Emulator boxes -- though we have not done the additional jackknife resampling step they did).

The thick solid and thin dashed curves in the top panel of Figure~\ref{fig:lag+std} show the cumulative distributions of $\alpha_{\rm LP}$ measured in the Laguerre reconstructed $\xi_{\rm Lag}$ and the standard density field reconstructions of the four HODs that were studied in the main text. This shows that although the standard reconstructions are slightly but not significantly biased (the median is not at $\alpha_{\rm LP} = 1$), the two reconstructions provide similar constraints on the LP-derived distance scale. This is consistent with Figures~\ref{fig:cdf} and ~\ref{fig:cdfPeak} in the main text.

The thick and thin solid curves in the middle panel of Figure~\ref{fig:lag+std} show the cumulative distributions of $\alpha_{\rm LP}$ and $\alpha_0$ for the Laguerre reconstructions $\xi_{\rm Lag}$ of the four HODs. The thin curves rise more steeply than the thick ones, indicating that fitting the full shape constrains the distance scale better than the LP does, by a factor of approximately 2. The cost, of course, is the dependence on fiducial template, and the potential biases which arise if the template is inaccurate.

The bottom panel of Figure~\ref{fig:lag+std} addresses the question of how well $\alpha_0$, estimated by fitting to $\xi_{\rm Lag}$ (thin solid), compares with that estimated from fitting to $\xi_0$ of the standard reconstruction (thin dot-dashed). The panel shows that the distributions are rather similar, indicating that even though $\xi_{\rm Lag}$ does not reconstruct the peak and dip scales very well (see Figure~\ref{fig:shape}), the fact that it is closer to linear theory over a wider range of scales mean its $\alpha_0$ values remain quite accurate. 

The main text notes that Laguerre reconstruction returns comparable precision and accuracy to standard density field reconstruction (for a small fraction of the computational cost). Together, the top and bottom panels show that this statement holds for both $\alpha_{\rm LP}$ and $\alpha_0$.


\begin{figure}
 \centering
 \includegraphics[width=0.9\hsize]{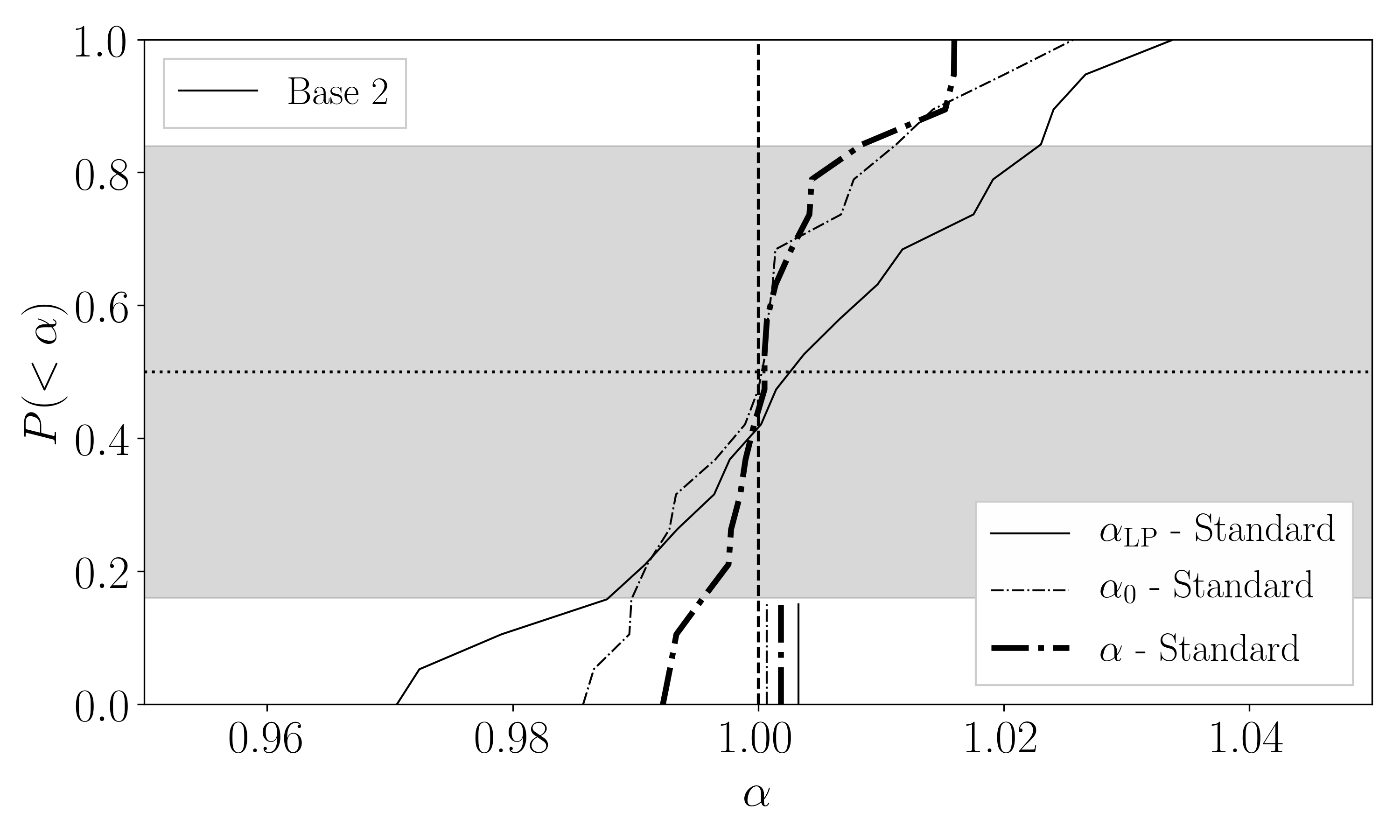}
 \caption{\rev{Cumulative distributions of $\alpha_{\rm LP}$ (solid), $\alpha_0$ (dashed, from fitting to $\xi_0$) and $\alpha\equiv(\alpha_\perp^2\alpha_{||})^{1/3}$ (dot-dashed, from fitting to both $\xi_0$ and $\xi_2$) in the standard reconstructions of the `Base 2' HOD.  The steepening of the curves illustrates the gains in precision as more assumptions about the expected shape of the correlation function are included.  Results for the other HODs we have considered are similar.}}
 \label{fig:alphas}
\end{figure}

We close with a direct comparison of how adding more information when estimating the distance scale can improve the constraints, provided that the template one fits to the data is accurate.  For this, we have chosen one of the HODs; the others make the same point.  The thin solid curve shows the distribution of $\alpha_{\rm LP}$ in the standard (density field) reconstruction, which makes {\em no} assumption about the shape of the correlation function.  The dashed curve shows the result of fitting to the monopole, $\xi_0$, of the redshift-space distorted correlation function in the same standard (density field) reconstruction, and the thick dot-dashed curve shows the result of fitting to both the monopole $\xi_0(s)$ and the quadrupole $\xi_2(s)$.  The distribution of $\alpha$ is slightly narrower than that of $\alpha_0$, showing that there are additional gains in accuracy if one fits to the full shape of $\xi(s_{||},s_\perp)$.  This provides strong motivation for extending the Laguerre methodology to the full redshift-space signal (rather than just the monopole).  Likewise, there is currently no analog of the linear point in $\xi_2$, nor, e.g., tests of its stability in angular wedges (other than those for which $\mu=0$, of course), so the difference between the thin solid curve and the other two motivates extending the linear point methodology to redshift space.

It is important to emphasize that the curves in Figures~\ref{fig:lag+std} and~\ref{fig:alphas} assume that the parameters needed for both reconstructions (Laguerre and standard) are known correctly, and that the template models which are needed to determine $\alpha_0$ or $\alpha$ are also known correctly.  In this respect, they are analogous to the optimistic `idealized' results in Section~\ref{sec:ideal} of the main text.  However, this will not be the case in observational datasets.  The results in Section~\ref{sec:realistic} of the main text illustrate how and why constraints from $\alpha_{\rm LP}$ are degraded when one marginalizes over uncertainties in the parameters needed for reconstruction.  As the main text discusses, obtaining more realistic error estimates for $\alpha_0$ and $\alpha$ is less straightforward.  This is because, to do so, one must also account for the fact that the template shape to which one should fit is not known perfectly.  Ref.~\cite{PRDforecast} argues that marginalizing over the full range of allowed template shapes will degrade the constraints from $\alpha_0$ and $\alpha$ significantly (a factor of two or more, given current uncertainties on cosmological parameters).  However, that analysis is based on pre-reconstruction quantities.  Ref.\cite{vm+2018} find that the impact of assuming an incorrect template ($\Omega_m$ wrong by 0.5 percent) when fitting to standard reconstructions impacts $\alpha$ at the level of 0.002, but a more exhaustive analysis, which marginalizes over the full range of allowed parameter space, and associated template shapes, has not yet been done.

\end{document}